\documentclass[journal,draftclsnofoot,onecolumn,12pt]{IEEEtran}
\IEEEoverridecommandlockouts

\usepackage{acronym}
\usepackage{amsfonts}
\usepackage[dvips]{graphicx}
\usepackage{times}
\usepackage{cite}
\usepackage{amsmath}
\usepackage{array}
\usepackage{amssymb}
\usepackage{stfloats}
\usepackage{diagbox}
\usepackage{graphicx}
\usepackage{footnote}
\usepackage{amsthm}
\usepackage{booktabs}
\usepackage{array}
\usepackage[ruled,vlined]{algorithm2e}
\usepackage{subeqnarray}
\usepackage{cases}
\usepackage{threeparttable}
\usepackage{color}
\usepackage{epstopdf}
\usepackage{multirow}
\usepackage{tabularx}
\usepackage{enumerate}
\usepackage{multicol}
\usepackage{hyperref}
\usepackage{subfig}
\usepackage{caption}

\def\bb0{{\mathbb{0}}}

\def\bb{{\mathbf{b}}}

\def\b0{{\mathbf{0}}}

\def\sf0{{\mathsf{0}}}

\def\rm0{{\mathrm{0}}}

\acrodef{CSI}[CSI]{channel state information}
\acrodef{CSIT}[CSIT]{channel state information at the transmitter}
\acrodef{CSIR}[CSIR]{channel state information at the receiver}
\acrodef{MIMO}[MIMO]{multiple-input multiple-output}
\acrodef{SISO}[SISO]{single-input single-output}
\acrodef{MISO}[MISO]{multiple-input single-output}
\acrodef{SIMO}[SIMO]{single-input multiple-output}
\acrodef{ADCs}[ADCs]{analog-to-digital convertors}
\acrodef{SNR}[SNR]{signal-to-noise ratio}
\acrodef{AWGN}[AWGN]{additive white Gaussian noise}
\acrodef{MRT}[MRT]{maximal ratio transmission}
\acrodef{DFT}[DFT]{Discrete Fourier Transform}
\acrodef{ULA}[ULA]{uniform linear array}
\acrodef{UPA}[UPA]{uniform planar array}
\acrodef{LS}[LS]{least squares}
\acrodef{ALMMSE}[ALMMSE]{approximate linear minimum mean squared error}
\acrodef{QIHT}[QIHT]{quantized iterative hard thresholding}
\acrodef{QIST}[QIST]{quantized iterative soft thresholding}
\acrodef{SVD}[SVD]{singular value decomposition}

\ifCLASSINFOpdf

\else
\fi
\begin{document}

\title{Cell-Free Massive MIMO for 6G Wireless Communication Networks}

\author{Hengtao He,~\IEEEmembership{Member,~IEEE,}
Xianghao Yu,~\IEEEmembership{Member,~IEEE,}
Jun Zhang,~\IEEEmembership{Senior Member,~IEEE,}
S.H. Song,~\IEEEmembership{Member,~IEEE,}
and Khaled B. Letaief,~\IEEEmembership{Fellow,~IEEE}

\thanks{H. He, X. Yu, J. Zhang and S. Song are with the Department of Electronic and Computer Engineering,
The Hong Kong University of Science and Technology, Hong Kong, E-mail: \{eehthe, eexyu, eejzhang, eeshsong\}@ust.hk.}
\thanks{Khaled B. Letaief is with the Department of Electronic and Computer Engineering, The Hong Kong University of Science and Technology, Hong Kong,
and also with Peng Cheng Laboratory, Shenzhen 518066, China (e-mail: eekhaled@ust.hk).}
}
\maketitle

\begin{abstract}
The recently commercialized fifth-generation (5G) wireless communication networks achieved many improvements, including air interface enhancement, spectrum expansion, and network intensification by several key technologies, such as massive multiple-input multiple-output (MIMO), millimeter-wave communications, and ultra-dense networking. Despite the deployment of 5G commercial systems, wireless communications is still facing many challenges to enable connected intelligence and a myriad of applications such as industrial Internet-of-things, autonomous systems, brain-computer interfaces, digital twin, tactile Internet, etc. Therefore, it is urgent to start research on the sixth-generation (6G) wireless communication systems. Among the candidate technologies for such systems, cell-free massive MIMO which combines the advantages of distributed systems and massive MIMO, is considered as a key solution to enhance the wireless transmission efficiency and  becomes the international frontier. In this paper, we present a comprehensive study on cell-free massive MIMO for 6G wireless communication networks, especially from the signal processing perspective. We focus on enabling physical layer technologies
for cell-free massive MIMO, such as user association, pilot assignment, transmitter and receiver design, as well as power control and allocation. Furthermore, some current and future research problems are highlighted.
\end{abstract}


\maketitle

\section{Introduction}\label{sec:intro}
The fifth-generation (5G) wireless communication networks have been deployed worldwide since 2019 to achieve  massive connectivity, ultra-reliability, and low latency. Among the several enabling technologies in 5G, massive multiple-input multiple-output (MIMO) \cite{massiveMIMO}, which deploys a large number of antennas at the base station (BS) in a {\em centralized} manner, can provide  very high beamforming and spatially multiplexing gain, thus achieving the high spectral efficiency (SE), energy efficiency, and link reliability. However, with the explosive demand for higher data rates and traffic volume, wireless communication networks are required to provide better coverage, and uniform user performance over a wide coverage area \cite{6G_khaled,6G_matthaiou,6G_you} which 5G cannot satisfy \cite{beyond5G}.  This is because the performance of massive MIMO systems, for instance, is restricted by the  inter-cell interference in the cellular network and the cell-edge users suffer significant performance degradation.

To improve the performance of cell-edge users, distributed antenna systems (DAS) have been proposed to cover the dead spots and offer  macro diversity in MIMO systems \cite{DAS}. It can provide better coverage and reduce the system power overhead. On the other hand,  Network MIMO and coordinated multi-point (CoMP) were proposed to reduce the inter-cell interference by adding cooperation between the neighboring access points (APs) \cite{Zhang, Network, CoMP}. They divide the APs into disjoint cooperation clusters to reduce the data sharing. However, interference between clusters is a critical issue because inter-cell interference cannot be removed within the cellular structure. This is because the inter-cell interference cannot be removed as long as the cellular paradigm is considered.

By combing the advantages of massive MIMO, DAS, and Network MIMO technologies, {\em cell-free massive MIMO} was proposed \cite{cell_free} where no cell and cell boundaries exists. Due to its natural advantages, cell-free massive MIMO has been regarded as a crucial and core technology for the upcoming sixth-generation (6G) networks. It is expected to bring important benefits, including huge data throughput, ultra-low latency, ultra-high reliability, high energy efficiency, and ubiquitous and uniform coverage \cite{6G_matthaiou,Ubiquitous}.
\begin{figure}
  \centering
  \includegraphics[width=5in]{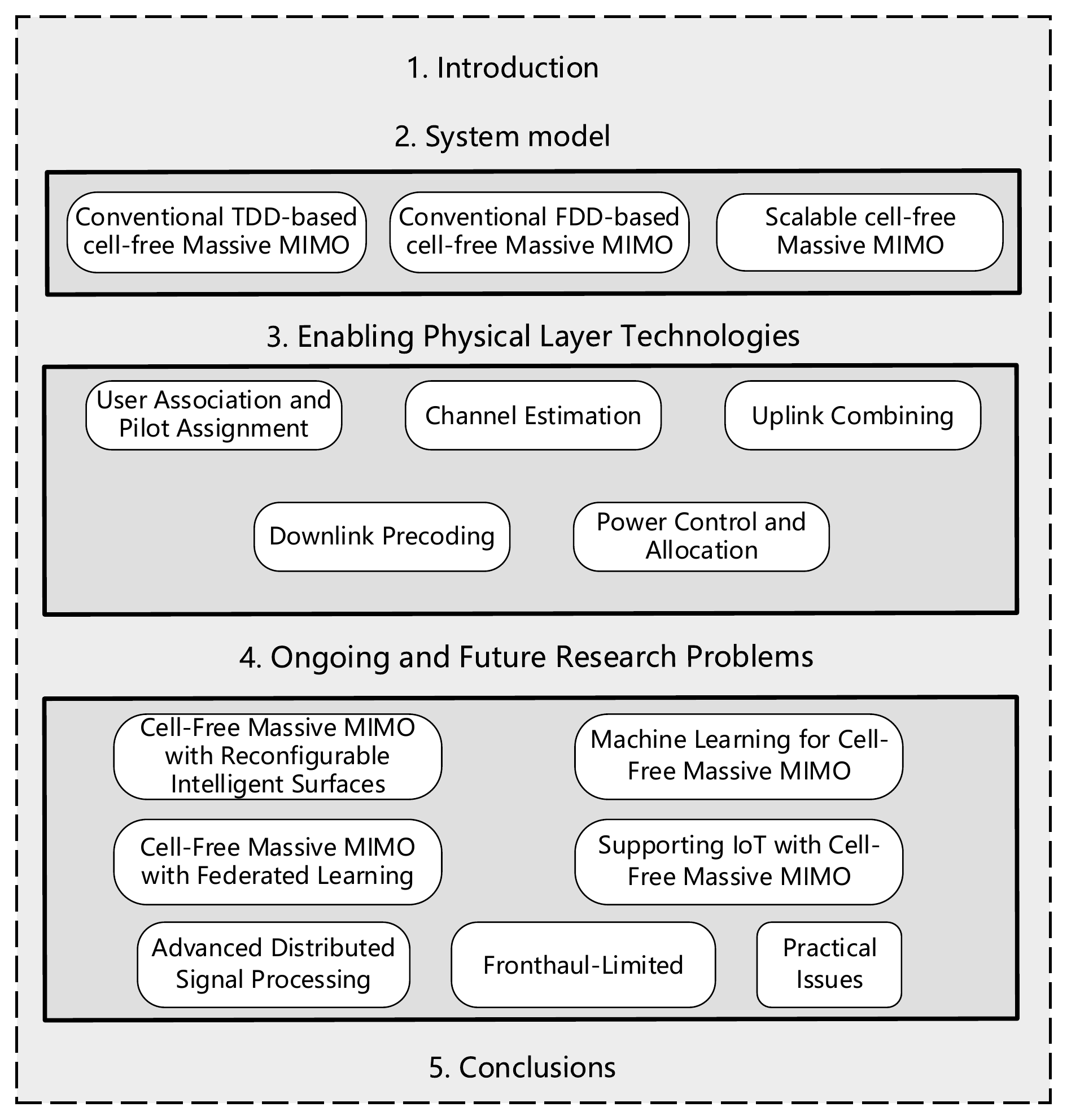}
  \caption{.~~Paper organization and main research directions of 6G cell-free massive MIMO.}\label{Fig:CF_block}
\end{figure}
The fundamental idea of cell-free massive MIMO is to deploy a large number of distributed  APs that are connected to a central processing unit (CPU) to serve all  users in a wide area. In particular, each AP serves all users via a time-division duplex (TDD) or frequency-division duplex (FDD) mode. Compared to conventional co-located massive MIMO, cell-free networks offer  more uniform connectivity for all users thanks to the macro diversity gain obtained from the distributed antennas. However, the assumption that each AP serves all users makes the system not scalable \cite{scalable} and incurs the huge power and computational resource consumption for decoding, especially for the users with low signal-to-interference-noise-ratios (SINR). To tackle the scalability issue, a \emph{user-centric} dynamic cooperation clustering (DCC) scheme \cite{DCC} was introduced \cite{scalable}, where each user is only served by a subset of APs. Therefore, the computational complexity and signaling overhead at each AP is finite even when the number of users goes to infinity \cite{scalable}.
\begin{table}[t!]
	\scriptsize
	\centering
	\begin{tabular}{|p{0.24\textwidth}|p{0.30\textwidth}|p{0.20\textwidth}|}
		\hline
		\hline
		\multicolumn{1}{|l|}{ \textit{\textbf{Topic}}}  & \multicolumn{1}{l|}{ \textit{\textbf{Proposed Solutions}}} & \multicolumn{1}{l|}{\textit{\textbf{Reference}}}\\
		\hline
		\hline

	\multirow{7}{*}{Pilot Assignment} & Random pilot assignment& \cite{random_pilot}
        \\
		\cline{2-3}
	 &	Location-based methods & \cite{local_based}
        \\
		\cline{2-3}
	 &	Graph coloring-based methods & \cite{graph_coloring}
        \\
		\cline{2-3}
     &  Hungarian algorithm & \cite{Pilot_Hungarian}
         \\
		\cline{2-3}
     &  Graph-based methods & \cite{Pilot_GNN}
		\\	
		\hline

        \multirow{2}{*}{Channel Estimation} & LMMSE  & \cite{decentalized_MMSE}
        \\
		\cline{2-3}
		 & DL-based methods & \cite{graph_coloring}	
		\\
        \hline
       \multirow{6}{*}{Uplink Combining} &  Level 1-4  & \cite{cell_free}
        \\
	    \cline{2-3}
	     & Sequential MMSE receiver & \cite{MMSE_optimal}	
	    \\
	    \cline{2-3}
	     & PM-based soft detection & \cite{PM}
        \\
	    \cline{2-3}
	     & EP-based detection & \cite{EP_distributed2}
        \\
		\cline{2-3}
		 & JCD &  \cite{JCD}
		\\
		\hline
        \multirow{6}{*}{Downlink Precoding}  & MRT precoding  & \cite{cell_free}
        \\
	    \cline{2-3}
	    & Enhanced MRT precoding & \cite{Enhanced_MRT}	
	    \\
	    \cline{2-3}
	    & ZF precoding & \cite{JMRZF}
        \\
	    \cline{2-3}
	    & Local ZF precoding & \cite{Local_ZF}
        \\
		\cline{2-3}
		& Team precoding &  \cite{Team_MMSE}
        \\
        \cline{2-3}
        & Cooperative precoding &  \cite{OTA_precoding}
        \\
		\hline
        \multirow{3}{*}{Power Control and Allocation}  & Optimization-based methods  & \cite{cell_free,JMRZF,SCA1,SCA2}
        \\
	    \cline{2-3}
	    & DL-based methods& \cite{power_control2, DL_Power_Con,DL_Power_JSAC}	
	    \\
	    \cline{2-3}
	    & Heuristic algorithms & \cite{power_control3}
        \\
		\hline
        \multirow{6}{*}{Ongoing and Future Problems}  & Cell-free with RIS  & \cite{Cell_free_RIS1,Cell_free_RIS2}
        \\
	    \cline{2-3}
	    & ML for cell-free & \cite{FFDet,DL_Power_JSAC,DL_Cascade}	
	    \\
	    \cline{2-3}
	    & Cell-free for FL & \cite{Cell_free_FL}
        \\
	    \cline{2-3}
	    & Advanced distributed signal processing & \cite{decentalized_MMSE,EP_distributed2}	
	    \\
	    \cline{2-3}
	    & Fronthaul-limited & \cite{quantization1,DL_fronthaul,quantization2,quantization3,quantization4}
	    \\
	    \cline{2-3}
	    & Practical issues & \cite{MMSE_optimal}
        \\
		\hline
	  \end{tabular}
	\caption{.~~Overview of different signal processing algorithms for cell-free massive MIMO systems}
	\label{table:overview}
\end{table}
Interesting research results have been obtained for cell-free massive MIMO, researchers made some initial attempts on analyzing the spectral and energy efficiency \cite{cell_free}, where single-antenna APs, single-antenna users, and Rayleigh fading channels are considered. The analysis has been extended to multi-antenna APs with Rayleigh fading, Rician fading, and correlated channels \cite{performance_analysis1,performance_analysis2,performance_analysis3}. The general conclusion is that cell-free massive MIMO works can achieve great performance in a variety of  scenarios. Then, the energy efficiency of cell-free massive MIMO systems was investigated \cite{performance_analysis6,performance_analysis8}. It was shown that the cell-free can improve the energy efficiency by approximately ten times compared to cellular massive MIMO. Therefore, cell-free massive MIMO has become one of the most promising technologies in 6G wireless networks and has been attracted extensive research interests from both academia and industry.
\begin{table}
\begin{tabular}{|p{0.1\textwidth}|p{0.74\textwidth}|}
  \hline
  \hline
  Symbols & Meanings \\
  \hline
  \hline
   $L$,$N$,$K$ & Number of APs, antennas in each AP, users \\
  \hline
  $\mathbf{h}_{kl}$ & The channel vector between $k$-th user and $l$-th AP \\
  \hline
  $\beta_{kl}$ & Large-scale fading \\
   \hline
  $\mathbf{R}_{kl}$ & Spatial correlation matrix \\
  \hline
  $\varphi_{t}$ & Pilot vector\\
  \hline
  $\tau_{p}$ & The length of the pilot\\
  \hline
  $\tau_{c}$ & Coherence time\\
  \hline
  $\mathcal{S}_{k}$ & The subset of users sharing the same pilot as the $k$-th user is denoted as \\
  \hline
  $t_{k}$ & The index of the pilot for $k$-th user\\
  \hline
  ${\bf Y}_{p}^{l}$ & Received signal corresponding to the pilot signal at the $l$-th AP\\
  \hline
  ${\bf N}_{p}^{l}$ & Additive noise at the $l$-th AP\\
  \hline
  $p_{k}$ & Additive noise at the $l$-th AP\\
  \hline
  $\sigma^{2}$ & Noise power\\
  \hline
  $s_{k}$ & Transmitted signal of $k$-th user\\
  \hline
  $\hat{s}_{k}$ & Estimated signal of $k$-th user\\
  \hline
  ${\bf y}_{d}^{l}$ & Received signal corresponding to the data at the $l$-th AP\\
  \hline
  $\mathbf{v}_{k}$ & Combining vector for $k$-th user\\
  \hline
  $\mathbf{D}_{kl}$ & Diagonal matrix to determine the connection between $l$-th AP and $k$-th user\\
  \hline
  $\mathbf{w}_{kl}$ & Precoding vector between $l$-th AP and $k$-th user\\
  \hline
  ${\bf n}$ & Additive noise vector\\
  \hline
  $\alpha_{p}$ & The complex gain of the $p$-th path\\
  \hline
  $\phi_{p}$ & The AoA of the $p$-th path\\
  \hline
  $\mathbf{a}(\phi_{p})$ & The steering vector\\
  \hline
  $\hat{\mathbf{h}}_{kl}$ & The channel estimate for the channel $k$-th user and $l$-th AP \\
  \hline
  $\tilde{\mathbf{h}}_{kl}$ & The channel estimation error for the channel $k$-th user and $l$-th AP \\
  \hline
  $\hat{\mathbf{R}}_{kl}$ & The spatial covariance matrix for the channel estimate $\hat{\mathbf{h}}_{kl}$ \\
  \hline
  $\tilde{\mathbf{R}}_{kl}$ & The spatial covariance matrix for the channel estimation error $\tilde{\mathbf{h}}_{kl}$ \\
  \hline
  $\mathcal{D}_{l}$ & The set of the user served by $l$-th AP \\
  \hline
\end{tabular}
	\caption{.~~Definition of the main mathematical symbols}
	\label{table:notations}
\end{table}
Although cell-free massive MIMO has shown a huge potential for 6G wireless networks, how to design effective algorithms for a low-cost and scalable system is significantly important. In this aspect, extensive algorithms have been proposed for solving specific problems in system design. In this paper, we provide a comprehensive overview on enabling physical layer technologies for cell-free massive MIMO, especially from the signal processing perspective. Different from other survey papers \cite{cell_free_Zhang,survey2,survey3} providing all research directions about cell-free massive MIMO in state-of-the-art literature, we pay attention to the existing signal processing algorithms for user association, pilot assignment, transmitter and receiver design, as well as power control and allocation. Furthermore, some future research directions, including machine learning (ML) and reconfigurable intelligent surfaces (RIS) for cell-free massive MIMO, are highlighted. The paper organization and main research directions are illustrated in Fig.\,\ref{Fig:CF_block}.

The remaining part of this paper is organized as follows. Section \ref{sec:system_model} first introduces the system model of  cell-free massive MIMO. Section \ref{physical} reviews existing signal processing algorithms in cell-free massive MIMO physical layer. 
Section \ref{sec:inter} discusses some open  directions in this area and our conclusions are given in Section \ref{sec:con}. The key topics and associated references are summarized in Table \ref{table:overview} and the definition of the main mathematical symbols is presented in Table \ref{table:notations}.
\section{System Model}\label{sec:system_model}
In this section, we introduce the system model of the cell-free massive MIMO, including conventional TDD-based cell-free massive MIMO, FDD-based cell-free massive MIMO, and scalable cell-free massive MIMO.
\subsection{Conventional TDD-based Cell-Free Massive MIMO}
As illustrated in Fig.\,\ref{cell_free}, a conventional cell-free massive MIMO systems consists of a large number of APs that  serve a much smaller
number of users on the same time-frequency resource. We consider a cell-free massive MIMO network consisting of $L$ distributed APs, each equipped with $N$ antennas to serve $K$ single-antenna users. All the APs are connected to a CPU. The network operates in the TDD or FDD mode and each AP acquires CSI between itself and all users via uplink channel estimation. The channel between the $l$-th AP and the $k$-th user is denoted by ${\bf h}_{kl}\in \mathbb{C}^{N}$ and is assumed to be constant in each coherence time $\tau_{c}$. The channel ${\bf h}_{kl}$ is assumed to be  a correlated Rayleigh fading distribution, i.e.,
\begin{equation}\label{eq_channel}
  {\bf h}_{kl}\sim \mathcal{N}_{\mathbb{C}}({\bf 0},{\bf R}_{kl}),
\end{equation}
where ${\bf R}_{kl}\in \mathbb{C}^{N \times N}$ denotes the spatial correlation matrix, which is composed of  the small-scale fading and large-scale fading. As such, the large-scale fading between the $k$-th user and the $l$-th AP is given by $\beta_{kl} \triangleq {\mathrm{tr}}({\bf R}_{kl})/N$, where ${\mathrm{tr}}(\mathbf{R}_{kl})$ denotes the trace of $\mathbf{\mathbf{R}}_{kl}$.

\begin{figure*}
\begin{minipage}{3in}
  \centerline{\includegraphics[width=3.0in]{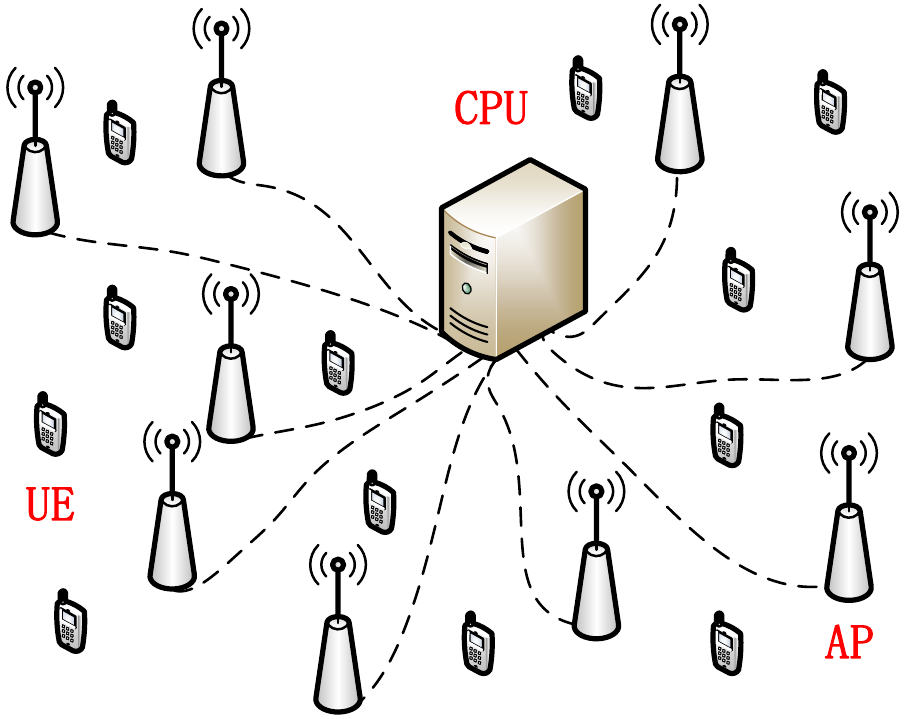}}
  \centerline{(a) conventional cell-free massive MIMO system}\label{cell_free}
\end{minipage}
\hfill
\begin{minipage}{3in}
  \centerline{\includegraphics[width=3.0in]{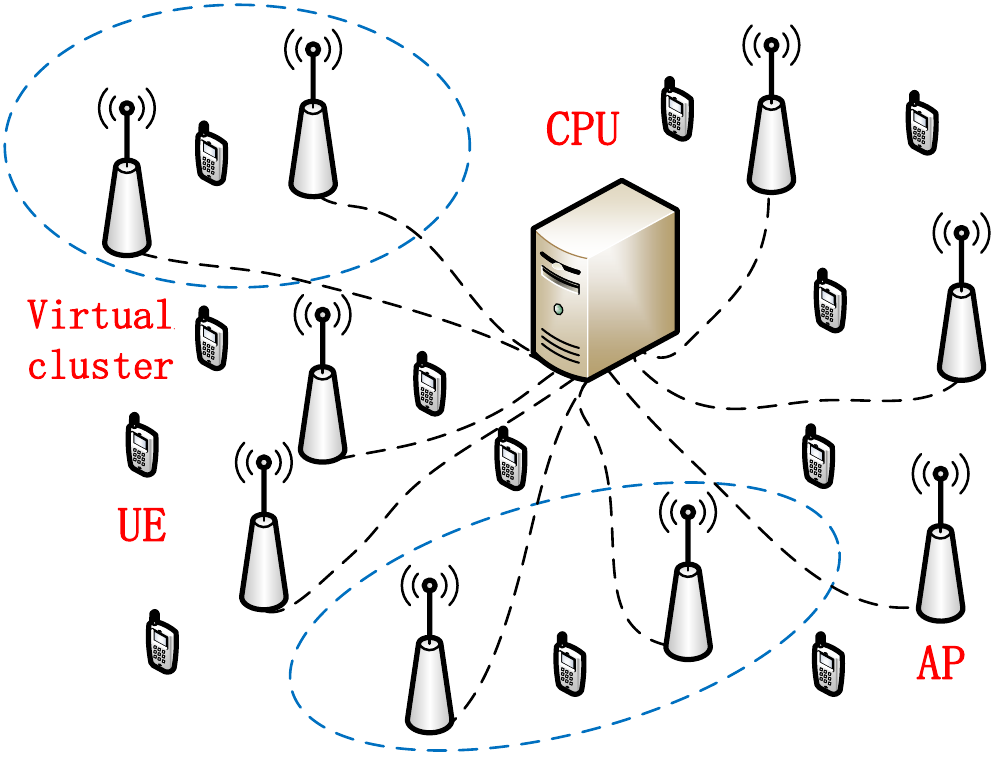}}
  \centerline{(b) scalable cell-free massive MIMO system.} \label{cell_free_DCC}
\end{minipage}
\caption{.~~Conventional and scalable cell-free massive MIMO system.}
\label{cell_free}
\end{figure*}

We assume that $\tau_{p}$ mutually orthogonal pilot $\varphi_{1},\ldots,\varphi_{\tau_{p}}$ with $\|\varphi_{t}\|^{2}_{2}=\tau_{p}, t = 1,\ldots,\tau_{p}$, are used for channel estimation, where $\tau_{p}$ is the length of the pilot. In general, the number of users $K>\tau_{p}$ and therefore each pilot is shared by more than one users. The index of the pilot assigned to the $k$-th user is denoted as $t_{k}\in \{1,\ldots,\tau_{p}\}$. Furthermore, the subset of users sharing the same pilot as the $k$-th user is denoted as $\mathcal{S}_{k}\subset \{1,\ldots, K\}$. Therefore, the uplink received signal corresponding to the pilot signal  at the $l$-th AP is
\begin{equation}\label{eqpilot}
  {\bf Y}_{l}^{p} = \sum_{i=1}^{K}\sqrt{p_{k}}{\bf h}_{kl}\varphi_{t_{k}}^{T}+{\bf N}_{l}^{p},
\end{equation}
where $p_{k}\geq 0$ represents the transmit power of the $k$-th user, ${\bf N}_{l}^{p}\in\mathbb{C}^{N\times\tau_{p}}$ denotes the additive noise and each element is independent and identically distributed (i.i.d.) and drawn from $\mathcal{N}_{\mathbb{C}}(0,\sigma^{2})$ with noise power $\sigma^{2}$. Based on estimation theory, the minimum mean-square error (MMSE) estimation for ${\bf h}_{kl}$ is given by
\begin{equation}\label{eqMMSE}
 \hat{{\bf h}}_{kl} =\sqrt{p_{k}\tau_{p}}\mathbf{R}_{kl}\Psi_{t_k l}^{-1}\mathbf{y}_{t_k l}^{p},
\end{equation}
where
\begin{equation}\label{eqCorrelation}
\Psi_{t_k l}= \sum_{i\in\mathbf{R}_{kl}}\tau_{p}p_{i}\mathbf{R}_{kl}+\mathbf{I}_{N},
\end{equation}
and
\begin{equation}\label{eqypilot}
\mathbf{y}_{t_k l}^{p} = {\bf Y}_{l}^{p}\frac{\varphi_{t_{k}}^{*}}{\sqrt{\tau_{p}}} =
\sum_{i\in\mathcal{S}_{k}}\sqrt{p_{k}\tau_{p}}{\bf h}_{il}+{\bf n}_{t_{k}l}.
\end{equation}

After channel estimation, the received signal ${\bf y}_{l}^{d}\in \mathbb{C}^{N}$ at the $l$-th AP in uplink data transmission stage is
\begin{equation}\label{eqyd}
   {\bf y}_{l}^{d} = \sum_{k=1}^{K}\sqrt{p_{k}}{\bf h}_{kl}s_{k}+{\bf n}_{l}^{d},
\end{equation}
where $s_{k}\in\mathbb{C}$ is the transmitted signal from the $k$-th user  with power $p_{k}$ and ${\bf n}^{d}_{l}\sim \mathcal{N}_{\mathbb{C}}(\mathbf{0},\sigma^{2}{\bf I}_{N})$. If the conventional cell-free massive MIMO considers the centralized decoding, the $l$-th AP first selects a receive combining vector ${\bf v}_{kl}\in \mathbb{C}^{N}$ for the $k$-th user and locally computes ${\bf v}_{kl}^{H}{\bf y}_{l}^{d}$. Then, the CPU  combines $s_{k}$ by following summation
\begin{align}\label{eqshat}
  \hat{s}_{k} & = \sum_{l=1}^{L}{\bf v}^{H}_{kl}{\bf y}_{l}^{p} \\
  & = \left(\sum_{l=1}^{L}  {\bf v}^{H}_{kl}{\bf h}_{kl}s_{k} \right) + \sum_{i=1,i\neq k}^{K}\left(\sum_{l=1}^{L}{\bf v}^{H}_{kl}{\bf h}_{kl}s_{i} \right)s_{i}+\sum_{l=1}^{L}{\bf v}^{H}_{kl}{\bf n}_{l} \\
  & = {\bf v}_{k}^{H}{\bf h}_{k}s_{k}+\sum_{i=1,i\neq k}^{K}{\bf v}_{k}^{H}{\bf h}_{i}s_{k}+{\bf v}^{H}_{k}{\bf n},
\end{align}
where ${\bf v}_{k} = [{\bf v}_{k1}^{T},\ldots,{\bf v}_{kL}^{T}]^{T} \in \mathbb{C}^{M}$ denotes the all combining vectors for of all APs and ${\bf n} = [{\bf n}_{1}^{T},\ldots,{\bf n}_{L}^{T}]^{T} \in \mathbb{C}^{M}$ collects all the noise vectors, where $M=LN$. The received signal model (\ref{eqshat}) is similar to a uplink cellular massive MIMO system model. Therefore, the achievable uplink spectral efficiency (SE) easily follow  related literature \cite{cell_free}. The key difference between cell-free and cellular massive MIMO is the combining vectors $\{{\bf v}_{kl}\}$ as each AP only have own CSI that estimated locally in the channel estimation stage. Furthermore, several combining methods have been proposed such as  maximum ratio combining (MRC) and local MMSE combining \cite{cell_free}.

\subsection{FDD-based Cell-Free Massive MIMO}
In addition to TDD systems, we consider an FDD cell-free massive MIMO system consisting of $L$ distributed APs, each equipped with $N$ antennas to serve $K$ single-antenna users, where the uplink and downlink transmissions are operated in different frequency bands. Owing to angle reciprocity\footnote{ Angle reciprocity means that the uplink and downlink channels have similar angles and complex gains.} in FDD systems, the uplink angle of arrival (AoA) and downlink angle of departure (AoD) are similar, and thus the uplink and downlink large-scale fading coefficients are similar  \cite{FDD_cell}. On the contrary, the uplink and downlink small-scale fading coefficients are different because they are frequency-dependent. Specially, the channel between the $k$-th user and $l$-th AP is given by
\begin{equation}\label{eqFDDchannel}
  \mathbf{h}_{kl} = \sqrt{\frac{1}{P}}\sum_{p=1}^{P}\sqrt{\beta^{kl}_{p}}\alpha_{p}\mathbf{a}(\phi_{p}),
\end{equation}
where $\alpha_{p}\sim \mathcal{N}_{\mathbb{C}}(0,1)$ is the complex gain of the $p$-th path, and $\beta^{kl}_{p}$ is the large-scale fading coefficient for the path-loss and shadowing effects. The AoA of the $p$-th path is $\phi_{p}\in[0,2\pi]$ and the steering vector $\mathbf{a}(\phi_{p})$ is given by
\begin{equation}\label{eqsteering}
 \mathbf{a}(\phi_{p}) = \frac{1}{\sqrt{N}}[1,e^{j d sin(\phi_{p})},\ldots,e^{j(N-1)d sin(\phi_{p})}],
\end{equation}
where $ d = \frac{2\pi u}{\lambda}$, $u$ is the antenna spacing, and $\lambda$ is the channel wavelength.

One of the challenges in FDD cell-free massive MIMO is the downlink CSI feedback. Because each AP receives the downlink CSI feedback from all users in conventional cell-free massive MIMO systems for designing the precoder, the CSI acquisition and feedback overhead will be huge. Fortunately, the angle reciprocity can be exploited if the uplink and downlink carrier frequencies are close to each other. In \cite{FDD_cell}, an FDD cell-free massive MIMO system without any feedback from the user was proposed, where the multipath components in the uplink stage can be extracted by a computationally efficient estimation algorithm  based on the gradient descent. The simulation results there showed that  FDD-based cell-free massive MIMO systems can reduce the transmission power when compared with conventional cellular systems.

\subsection{Scalable Cell-Free Massive MIMO}
Although the centralized processing in conventional cell-free massive MIMO can achieve great performance, it is impractical when the number of users is large. According to \cite{scalable}, a cell-free massive MIMO is considered to be scalable if the signal processing tasks for channel estimation, precoder and combiner design, fronthaul overhead, and power optimization per AP can be kept within finite complexity as the number of served users goes to infinity. Unfortunately, the conventional cell-free massive MIMO is not scalable with respect to all of the four tasks as shown below:

1) Precoder and combiner design: The $l$-th AP computes precoding and combining vectors for all $K$ users. The computational complexity increases to infinity as $K \rightarrow \infty$.

2) Channel estimation: As the $l$-th AP performs channel estimation for all $K$ users, the computational complexity increases to infinity as $K \rightarrow \infty$.

3) Fronthaul overhead: As the $l$-th AP needs to recover $K$ downlink data signals and forward $K$ received signals ${\bf v}_{kl}^{H}{\bf y}_{l}^{p}$ to the CPU.
The Fronthaul overhead increases to infinity as $K \rightarrow \infty$.

4) Power optimization: If centralized power optimization is considered, the computational complexity increases to infinity as $K \rightarrow \infty$.

The aforementioned scalability issues can be solved by constructing scalable cell-free massive MIMO network with the DCC framework \cite{scalable}. We utilize a set of diagonal matrices ${\bf D}_{kl}\in \mathbb{C}^{N \times N}$, for $k = 1,\ldots, K$ and $l = 1,\ldots, L$, to determine the connection between antennas and users. Specifically,  ${\bf D}_{kl}$ is an identity matrix if it is allowed to serve the $k$-th user. The DCC framework changes the received uplink signals ${\bf y}_{l}^{d}$. Specifically, only a subset of the APs need to participate the signal detection, and the estimate of data can be expressed as
\begin{align}\label{eqshat2}
  \hat{s}_{k} & = \sum_{l=1}^{L}{\bf v}^{H}_{kl}{\bf D}_{kl}{\bf y}_{l}^{d} \\
  & = {\bf v}_{k}^{H}{\bf D}_{k}{\bf h}_{k}s_{k}+\sum_{i=1,i\neq k}^{K}{\bf v}_{k}^{H}{\bf D}_{k}{\bf h}_{i}s_{i}+{\bf v}^{H}_{k}{\bf D}_{k}{\bf n}_{k},
\end{align}
where ${\bf D}_{k} = \mathrm{diag}({\bf D}_{k1},\ldots,{\bf D}_{kL}) \in \mathbb{C}^{M \times M}$ is a block-diagonal matrix. Similarly, the received downlink signal at $k$-th user is given by
\begin{equation}\label{eqDCC}
  y_{k}^{dl} = \sum\limits_{l=1}^{L}{\bf h}_{kl}^{H}\sum\limits_{k=1}^{K}{\bf D}_{kl}{\bf w}_{kl}q_{k}+n_{k}.
\end{equation}
where ${\bf w}_{kl}$ is the precoding vector, $q_{k}$ is the transmitted signal, and $n_{k}\sim \mathcal{N}_{\mathbb{C}}(0,\sigma^{2})$ is the additive Gaussian noise. If ${\bf D}_{kl}=\mathbf{0}$,  the $k$-th user is not served by $l$-th AP. Thus, the APs with ${\bf D}_{kl} \neq \mathbf{0}$  transmit to the $k$-th user in the downlink and apply receive combining in the uplink. Note that the conventional cell-free massive MIMO can be treated as the special case with ${\bf D}_{kl}={\bf I}_{N}$ for $i = 1,\ldots, K$ and $l = 1,\ldots, L$, where all antennas in each AP serve all users.
\section{Enabling Physical Layer Technologies for Cell-Free Massive MIMO}\label{physical}
In this section, we discuss and present key enabling physical layer technologies for cell-free massive MIMO. Specially, we focus on
state-of-the-art signal processing algorithms for several challenging problems, including user association, pilot assignment, channel estimation, uplink combining, downlink precoding, and power control and allocation.
\subsection{User Association and Pilot Assignment}\label{sec:user_ass}
If a user is required to be served in the network, the system should assign a pilot first and construct a set $\mathcal{D}_{l}$ for each pilot. The conventional cell-free massive MIMO architecture assumes that each AP serves all users in the network simultaneously, which is impractical in real systems. With scalable cell-free systems, each AP only serves several users owing to the pilot shortage and avoids strong pilot contamination. Therefore, each user is served by a subset of APs. Therefore, user association and pilot assignment are great of importance in scalable cell-free massive MIMO systems.

In \cite{scalable}, a three-step association procedure was proposed for joint initial association, pilot assignment, and cluster formation. Assume that the $k$-th user is required to be served, it first measures the large-scale fading factor
$\beta_{kl} = \mathrm{tr}({\bf R}_{kl})/N$ for all near APs. Then, the $k$-th user appoints AP with
\begin{equation}\label{eqlarge}
  l = \mathrm{argmax}_{l} \beta_{kl},
\end{equation}
as its master AP. Subsequently, the appointed master AP assigns the $\tau$-th pilot  to the user, where
\begin{equation}\label{eqtau}
  \tau = \mathrm{argmin}_{t} \mathrm{tr}(\Psi_{tl}),
\end{equation}
with $\Psi_{tl}$ is shown in (\ref{eqCorrelation}). Above-mentioned process operates among the existing ($k-1$) users and assign the pilot $\tau$ to the $k$-th user. Then, the master AP  informs several neighboring about the pilot information. Afterwards, each neighboring AP decides whether to serve the $k$-th user based on some rules. If a user moves around or other users leave or connect to the network, the association procedure is required to be operated again.

Furthermore, by maximizing the lower bound for the uplink SE within the subset of the APs, the association problem is formulated as a matching problem as follows,
\begin{align}\label{eqmatch}
 & \max \limits_{z_{k,j}}\sum\limits_{k=1}^{K}\sum\limits_{j=1}^{J}z_{k,j} log_{2}(1+\mathrm{SINR}_{k}^{(j)}) \\
  &  \mathrm{s.t.} \sum\limits_{k=1}^{K}z_{k,j}=1,\quad \forall j =1,...,J \\ \nonumber
   &   z_{k,j} \in \{0,1\},\quad \forall k,j, \nonumber
\end{align}
where $\mathrm{SINR}_{k}^{(j)}$ is the SNR expression for the $k$-th user connected with $j$-th virtual cluster. $z_{k,j}$ indicates  whether the $k$-th user is served by the $j$-th virtual cluster. Specifically,  $z_{k,j}=1$ if the $k$-th user is assigned to the $j$-th virtual cluster. The matching problem illustrated in (\ref{eqmatch}) can be addressed by the Hungarian algorithm in polynomial time, which is known as an efficient  combinatorial algorithms for solving weighted matching problem \cite{user_association}. Furthermore, only the knowledge of the position of the APs  are needed for the Hungarian algorithm. Although numerical results show that the Hungarian algorithm is not always better than other algorithms, it can achieve a lower backhaul overhead and scalable system with a marginal performance loss compared to conventional cell-free massive MIMO.

CSI acquisition is significantly important for cell-free massive MIMO to achieve the potential performance. Owing to the large number of users served by the system and limited length of the orthogonal pilot, pilot reuse is inevitable. However, pilot contamination  will deteriorate the system performance. In \cite{cell_free}, the  random pilot assignment scheme was considered to allow each user  randomly choose a pilot sequence which causes the sever pilot contamination. Then, a pilot assignment scheme based on greedy algorithm was proposed in \cite{cell_free} to update the lowest uplink achievable rate of all users iteratively. However, it  only improves the worst user's performance instead of the total system performance.  Subsequently, to maximize the minimum distance among users with the reused pilots, a structured pilot assignment scheme was proposed.

The location information has been utilized into pilot assignment before using the greedy pilot assignment algorithm \cite{local_based}. However, it can only promote limited throughput performance. In \cite{graph_coloring}, a graph coloring-based  pilot assignment scheme   was proposed to significantly reduce the pilot contamination. Note that a common intention in \cite{local_based,graph_coloring} is to formulate appropriate policies to avoid pilot reuse among nearby users. However, it is not sufficiently accurate to measure pilot contamination  only by geographical proximity between users. Then, the Hungarian algorithm was used to for pilot assignment \cite{Pilot_Hungarian} and provides more performance improvement when the number of orthogonal pilot is significantly smaller than the number of users. Furthermore, a novel pilot assignment scheme based on a weighted graphic framework  was proposed  to alleviate the pilot contamination problem and reinforce the quality of service (QoS) in cell-free massive MIMO systems \cite{Pilot_GNN}.

The above pilot assignment based on the assumption that the pilot training and data transmission are performed separately in each coherence time. On the contrary, the superimposed pilots, which achieve the simultaneous transmission of the pilot and data symbols in each coherence time, can be utilized in cell-free massive MIMO systems. In \cite{Superimposed}, the achievable SE analysis of an uplink cell-free massive MIMO system with superimposed pilots-aided channel estimation was investigated. It outperforms the conventional pilot training and data transmission approaches in terms of the channel estimation performance and achievable rate.

\subsection{Channel Estimation}\label{sec:channel_estimation}
To achieve the potential performance of cell-free massive MIMO, accurate CSI should be obtained at the CPU or APs. For TDD system, channel reciprocity can be exploited to obtain the downlink CSI. The uplink received signal corresponding to pilot signal at the $l$-th AP is
\begin{equation}\label{eqpilot}
  {\bf Y}_{l}^{p} = \sum_{i=1}^{K}\sqrt{p_{i}}{\bf h}_{il}\varphi_{t_{i}}^{T}+{\bf N}_{l}^{p},
\end{equation}
where $p_{i}\geq 0$ is the transmit power of the $i$-th user , ${\bf N}_{l}^{p}\in\mathbb{C}^{N\times\tau_{p}}$ is the additive noise and each element is drawn from $\mathcal{N}_{\mathbb{C}}(0,\sigma^{2})$ with noise power $\sigma^{2}$. Based on estimation theory, the MMSE estimate of ${\bf h}_{kl}$ is
\begin{equation}\label{eqMMSE}
 \hat{{\bf h}}_{kl} =\sqrt{p_{k}\tau_{p}}\mathbf{R}_{kl}\Psi_{t_k l}^{-1}\mathbf{y}_{t_k l}^{p},
\end{equation}
where
\begin{equation}\label{eqypilot}
\mathbf{y}_{t_k l}^{p} = {\bf y}_{l}^{p}\frac{\varphi_{t_{k}}^{*}}{\sqrt{\tau_{p}}} =
\sum_{i\in\mathcal{S}_{k}}\sqrt{p_{k}\tau_{p}}{\bf h}_{il}+{\bf n}_{t_{k}l},
\end{equation}
and $\mathbf{y}_{t_k l}^{p}$ and  $\Psi_{t_k l}$ are the received signal and its covariance matrix, respectively. ${\bf n}_{t_{k}l} \triangleq {\bf N}_{l}\varphi_{t_{k}}^{*}/\sqrt{\tau_{p}}\sim \mathcal{N}_{\mathbb{C}}(\mathbf{0},\sigma^{2}\mathbf{I}_{N})$ is the equivalent noise. An important characteristic of MMSE estimation is the that  the estimate ${\bf h}_{kl}\sim \mathcal{N}_{\mathbb{C}}(\mathbf{0},\hat{\mathbf{R}}_{kl})$ and the estimation error $
\tilde{{\bf h}}_{kl} = {\bf h}_{kl}-\hat{{\bf h}}_{kl}\sim \mathcal{N}_{\mathbb{C}}(\mathbf{0},\tilde{\mathbf{R}}_{kl})$ is statistical independent. The  covariance matrix are
\begin{align}\label{eqRhat}
\hat{{\bf R}}_{kl} &= \mathbb{E}\left\{ (\hat{{\bf h}}_{kl}- \mathbb{E}\{\hat{{\bf h}}_{kl}\})(\hat{{\bf h}}_{kl}- \mathbb{E}\{\hat{{\bf h}}_{kl}\})^{H} \right\}  \\
& = p_{k}\tau_{p}\hat{{\bf R}}_{kl}\Psi_{t_k l}^{-1}\hat{{\bf R}}_{kl},
\end{align}
\begin{align}\label{eqtilde}
\tilde{{\bf R}}_{kl} &= \mathbb{E}\left\{ (\tilde{{\bf h}}_{kl}- \mathbb{E}\{\tilde{{\bf h}}_{kl}\})(\tilde{{\bf h}}_{kl}- \mathbb{E}\{\tilde{{\bf h}}_{kl}\})^{H} \right\}  \\
& = p_{k}\tau_{p}\tilde{{\bf R}}_{kl}\Psi_{t_k l}^{-1}\tilde{{\bf R}}_{kl},
\end{align}
respectively.

Channel estimation can be divided according to different levels of centralization. For fully centralized channel estimation, the pilot signals received by the APs are transmitted to the CPU, then the CPU performs channel estimation. Another way is to perform channel estimation at the AP locally to estimate the channels of its associated users. If Rayleigh fading channel is considered, MMSE estimator is often applied in the channel estimation stage \cite{decentalized_MMSE}. Furthermore, the mmWave channel has been considered in cell-free massive MIMO,  and compressed sensing (CS) or DL-based channel estimator has been proposed  accordingly \cite{FFDet}. The fast and flexible denoising convolutional neural network proposed in \cite{FFDet}  extracts the mmWave channel to achieve excellent channel estimation performance, and outperforms several CS-based channel estimator. Different from co-located massive MIMO where the users do not need to estimate the downlink channel by virtue of channel hardening, the cell-free networks provide a low degree of channel hardening. Therefore, the downlink beamforming was investigated to improve the achievable downlink rate, although pilot overhead and additional pilot contamination are introduced.

When centralized channel estimation is considered, users' location information should be known at the CPU, which causes the leakage of the user's privacy. To avoid disclosing the users' location and privacy, differential privacy (DP)-based channel estimation algorithms were proposed for cell-free hybrid massive MIMO system \cite{DP_channel_estimation}. Two privacy-preserving channel estimators based on Frank-Wolfe iteration and singular value decomposition were investigated, respectively, and the estimation error bounds for the two algorithms were analyzed.

FDD-based cell-free massive MIMO was also investigated in \cite{FDD_cell}. The key idea is to exploit the angle-reciprocity of the multi-path channel and the CSI acquisition overhead is reduced accordingly.  Downlink CSI estimation can further  benefit from angle reciprocity by uplink-aided downlink channel estimation when the uplink and downlink carrier frequencies are relatively close to each other. A new feedback reduction technique that exploits angle reciprocity was proposed for cell-free massive MIMO. Only the  information of a few selected dominant paths is feedback to BS and  the angle information is obtained by uplink pilot signal.

\subsection{Uplink Combining}\label{sec:Uplink_Combining}
Efficient data detection algorithms are highly desired in large-scale and complex networks, such as cell-free massive MIMO systems. For uplink detection,  the transmitted data ${\bf x}$ is estimated based on the received signals ${\bf y}_{l}\,(l=1,2,\ldots,L)$, channel matrix ${\bf H}$, and noise power $\sigma^{2}$. In this aspect, some early attempts were made on centralized algorithms where the detection is performed at the CPU with the received pilots and data signals reported from all APs \cite{cell_free, LSFD}. However, the computational overhead of such a centralized detection scheme is prohibitively high when the network size becomes large. In \cite{decentalized_MMSE}, one centralized and three distributed receivers with different levels of cooperation among APs were compared in terms of SE. These levels are defined as follows:

\begin{itemize}
  \item Level $4$ is a fully centralized receiver where the pilot and data signals received at all APs are sent to the CPU for channel estimation and data detection. In this case, the MMSE estimator for each user is given by
\begin{equation}\label{eqcombiner4}
{\bf v}_{k} =p_{k}\bigg(\sum\limits_{i=1}^{K}p_{i}(\hat{{\bf h}_{i}}\hat{{\bf h}_{i}^{H}}+\tilde{{\bf R}}_{i}) + \sigma^{2}{\bf I}_{LN}\bigg)^{-1}\hat{{\bf h}}_{k}
\end{equation}

\item Level $3$ involves two stages. First, each AP estimates the channels and uses the linear MMSE detector to detect the received signals. Then, the detected signals are collected at the CPU for joint detection for all UEs by utilizing the large-scale fading decoding (LSFD) method. Compared to Level $4$, only the channel statistics are utilized at the CPU but the pilot signals are not required to be sent to the CPU. The MMSE combining vector is given by,
\begin{equation}\label{eqcombiner3}
{\bf v}_{kl} =p_{k}\bigg(\sum\limits_{i=1}^{K}p_{i}(\hat{{\bf h}_{il}}\hat{{\bf h}_{il}^{H}}+\tilde{{\bf R}}_{il}) + \sigma^{2}{\bf I}_{LN}\bigg)^{-1}\hat{{\bf h}}_{k}
\end{equation}

and the local estimate is given by $\check{s}_{kl} = {\bf v}_{kl}^{H}{\bf y}_{l}$. The local estimate $\check{s}_{kl}\,(l=1,2,\ldots,L)$  is then sent to the CPU where combined by using the weights $a_{kl}\,(l=1,2,\ldots,L)$ to obtain $\hat{s}_{k} =\sum_{l=1}^{L} a^{*}_{kl}\check{s}_{kl}$. The optimal
vector ${\bf a}_{k}$ to maximize the SINR for each user is given by,

\begin{equation}\label{eqak}
{\bf a}_{k} = \bigg(\sum\limits_{i=1}^{K}p_{i} \mathbb{E}\{{\bf g}_{ki}{\bf g}_{ki}^{H}\}+\sigma^{2}{\bf E}_{k}                               \bigg)^{-1}\mathbb{E}\{{\bf g}_{kk}\}
\end{equation}
where ${\bf a}_{k} = [a_{k1}, \ldots, a_{kL} ]^{T} \in \mathbb{C}^{L}$ is the weighting coefficient vector,
${\bf g}_{ki} = [{\bf v}_{k1}^{H}{\bf h}_{i1}, \ldots, {\bf v}_{kL}^{H}{\bf h}_{iL}]$  is the vector with respect to the receive-combined channels between the $k$-th user and each AP, and ${\bf E}_{k}  = \mathrm{diag}(\mathbb{E}\{\|\mathbf{v}_{k1}\|^{2}\},\ldots,\mathbb{E}\{\|\mathbf{v}_{kL}\|^{2}\})\in \mathbb{C}^{L\times L}$, respectively.

\item Level $2$ is a special case of Level $3$. The CPU performs joint detection for all UEs by simply taking the average of the local estimates. Thus, no channel statistics is required to be transmitted to CPU via the fronthaul and the estimated signal in CPU is given by,

\begin{equation}\label{eqak}
   \hat{s}_{k} = \frac{1}{L} \sum\limits_{l=1}^{L}\check{s}_{kl},
\end{equation}
where  $\check{s}_{kl} = {\bf v}_{kl}^{H}{\bf y}_{l}$. When ${\bf a}_{k} = [1/L \ldots 1/L]^{T}$, Level 3 is reduced to Level 2.

\item Level $1$ is a fully distributed approach where the data detection is performed  at the APs based on the local channel estimates. No information is required to be transferred to the CPU.
\end{itemize}

Although the centralized processing is more attractive in terms of SE and fronthaul overhead, the system is not scalable if the number of APs and users is large. Furthermore, the aforementioned distributed receiver require large fronthaul capacity between the APs and the CPU, which is difficult for practical systems. To solve the problem, the radio stripes were incorporated into cell-free massive MIMO in \cite{decentalized_MMSE}. The APs are sequentially connected and share the same fronthaul link in a radio stripe network,  which reduces the cabling substantially. Based on the structure, a novel uplink sequential processing algorithm was developed which can achieve the optimal performance in terms of both SE and MSE. It can achieve the same performance as the centralized MMSE processing, but requires much lower fronthaul overhead and makes full use of the computational resource at the APs.

An alternative way is to consider the non-linear detector for cell-free massive MIMO. As linear detection performed at the AP or CPU is highly suboptimal and could even be ill-conditioned, a potential solution is to design a distributed non-linear detector for cell-free network. In \cite{soft_detection}, a partial marginalization (PM)-based  soft detection was proposed \cite{PM}, where each AP locally implements the non-linear detection and shares the so-obtained per-bit log-likelihood ratios on the fronthaul link. The data decoding is performed at the CPU by collecting soft bits from the APs for each user and each AP only decodes a subset of users.

Joint channel estimation and data detection (JCD), which exploits the channel sparsity and finite-alphabet constellation, have been investigated for cell-free massive MIMO \cite{JCD}. It first formulates the JCD as a biconvex optimization problem given by,
\begin{equation}\label{eqak}
 \{ \hat{{\bf H}}, \hat{{\bf S}}_{D} \} = \arg\min_{\mathbf{H},\mathbf{S}_{D}}\frac{1}{2}\| \mathbf{Y}-\mathbf{H}[\mathbf{S}_{T},\mathbf{S}_{D}]    \|_{F}^{2}+\mu\|\mathbf{H}\|_{1},
\end{equation}

where $\mathbf{S}_{T}$ and $\mathbf{S}_{D}$ are the pilot and data signal, respectively. The forward-backward splitting algorithm is adopted to solve the problem. The proposed JCD algorithm can support cell-free systems where the number of users is comparable to the number of AP, and utilizes the non-orthogonal training sequence to reduce the training overhead.

\begin{figure}[t]
  \centering
  \includegraphics[width=3.5in]{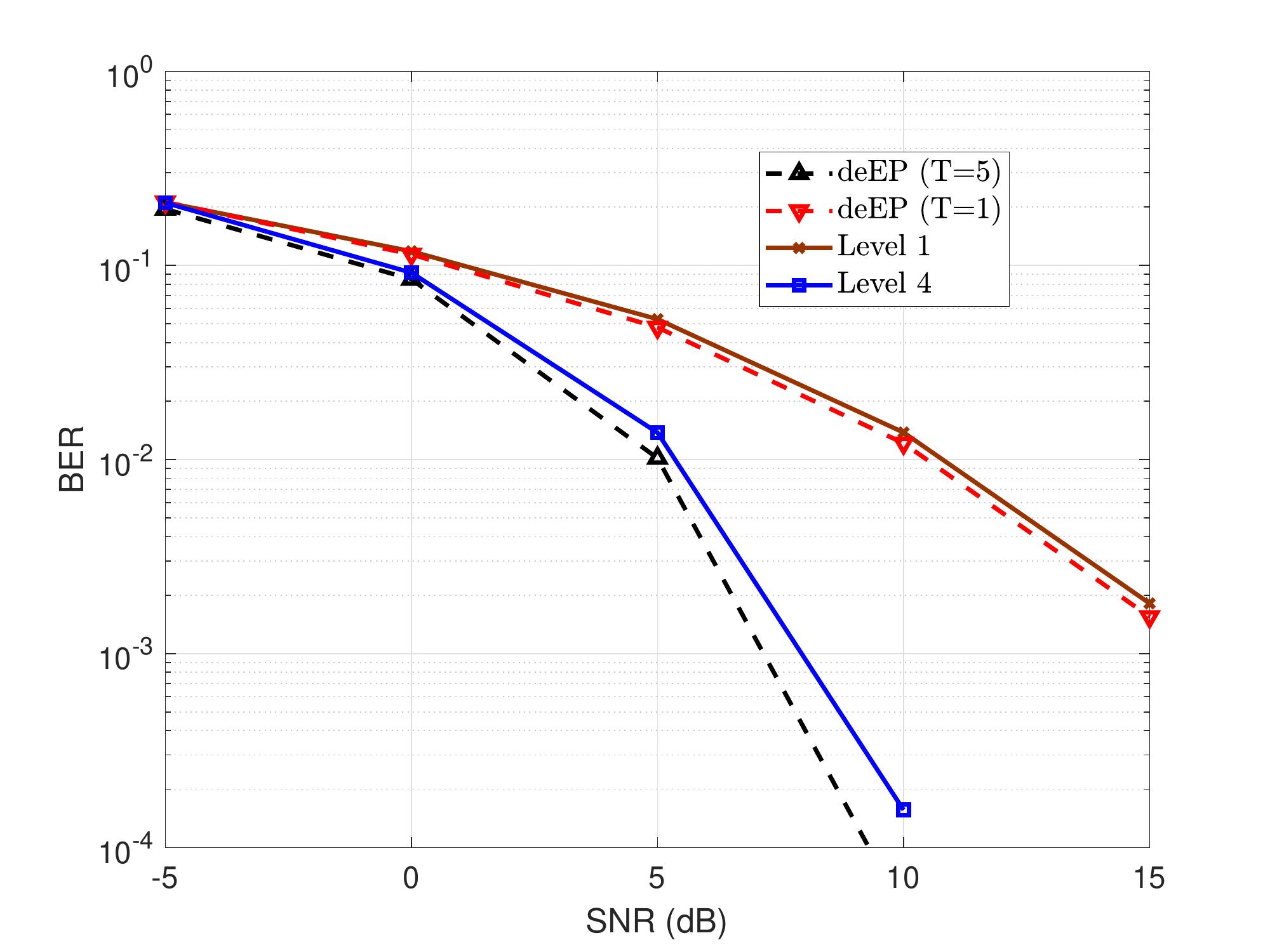}
  \caption{.~~BER performance comparisons of different detectors in conventional cell-free massive MIMO systems.}\label{Fig:nonscalable}
\end{figure}
From Bayes' theorem,  the Bayesian MMSE data detector is an optimal detector in MMSE sense, where the posterior probability density function is given by
\begin{equation}\label{eq4}
    \mathtt{P}(\mathbf{x}|\mathbf{y},\mathbf{H})=\frac{\mathtt{P}(\mathbf{y}|\mathbf{x},\mathbf{H})\mathtt{P}(\mathbf{x})}{\mathtt{P}(\mathbf{y}|\mathbf{H})}
 = \frac{\mathtt{P}(\mathbf{y}|\mathbf{x},\mathbf{H})\mathtt{P}(\mathbf{x})}{\int \mathtt{P}(\mathbf{y}|\mathbf{x},\mathbf{H})\mathtt{P}(\mathbf{x})d\mathbf{x}}.
\end{equation}

Given the posterior probability density function $\mathtt{P}(\mathbf{x}|\mathbf{y},\mathbf{H})$, the Bayesian MMSE estimate is obtained by
\begin{equation}\label{MMSE_estimate}
\hat{\mathbf{x}}=\int \mathbf{x} \mathtt{P}(\mathbf{x}|\mathbf{y},\mathbf{H})d\mathbf{x}.
\end{equation}
However, the Bayesian MMSE estimator is not computationally tractable because the marginal posterior probability in (\ref{MMSE_estimate}) involves a high-dimensional integral. The expectation propagation (EP) algorithm, proposed in \cite{EP_principle}, provides an iterative method to recover the transmitted $\mathbf{x}$ from the received signal $\mathbf{y}$ and has been recently attracting extensive research interests \cite{EP_principle,Decentralized-AMP,EP_distributed,EP_distributed2,Takeuchi,EP_detector,HeJSTSP}.
It is derived from the factor graph with the messages updated and passed between different pairs of nodes assumed to follow Gaussian distributions. Its distributed variants can be applied into cell-free massive MIMO systems and was shown to be able to achieve excellent performance \cite{EP_distributed2}.
\begin{figure}
  \centering
  \includegraphics[width=3.5in]{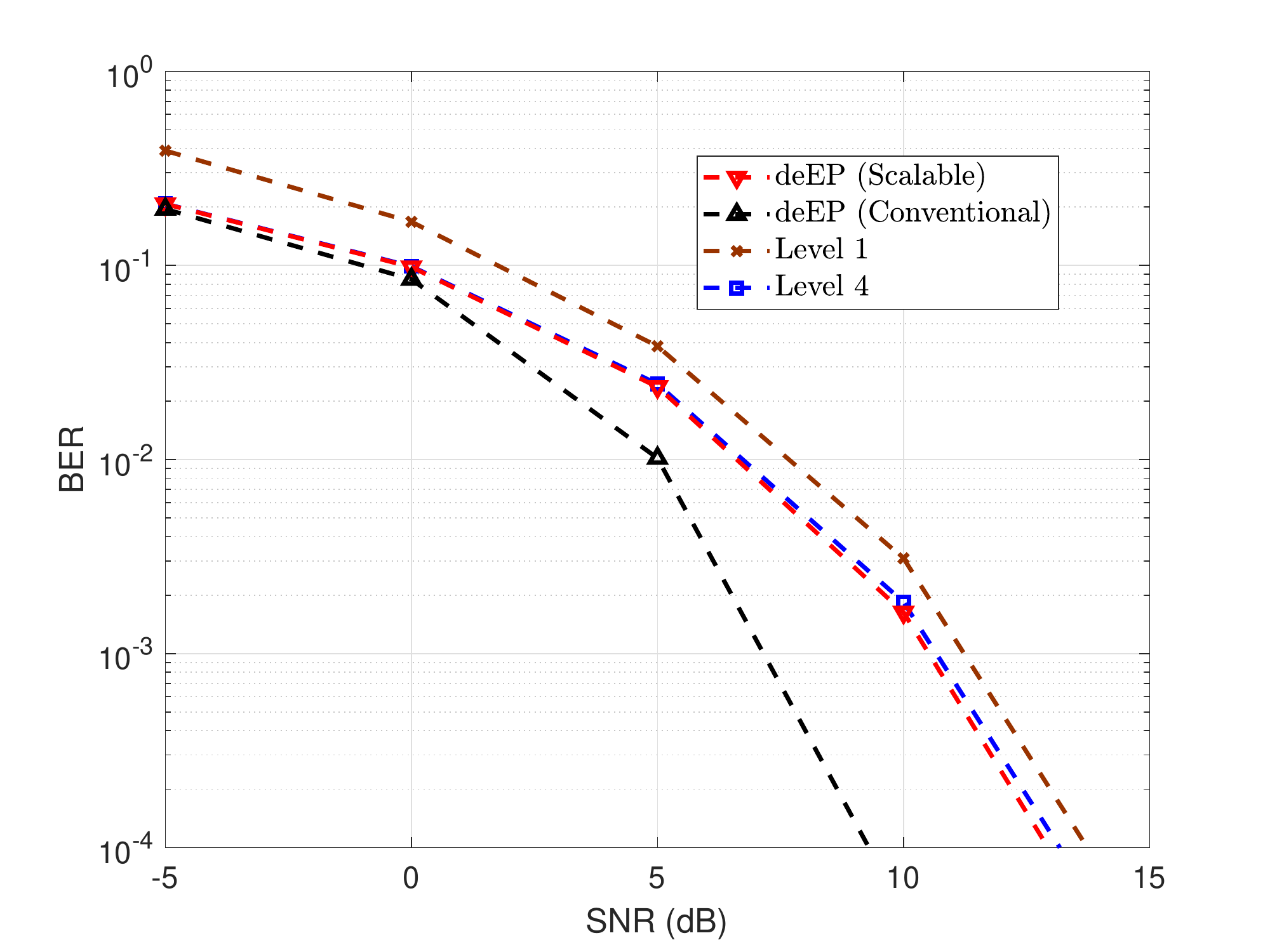}
  \caption{.~~BER performance comparisons of different detectors in scalable cell-free massive MIMO systems.}\label{Fig:DCC}
\end{figure}

\begin{footnotesize}
\begin{table}[t]
	\centering
	\renewcommand{\arraystretch}{1.5}
	\begin{minipage}[c]{1\columnwidth}
		\centering
		\caption{.~~Complexity of different detectors}
		\label{tbl:complexity}
		\begin{tabular}{@{}lcccc@{}}
			\toprule
			Detectors & Level 1 & Level 4  & deEP   \\
	    	\midrule
              CPU & $0$ & $O(LN)^{3}$&$O(T|\mathcal{D}_{l}|^{2})$ \\
			\midrule
			  AP & $O(|\mathcal{D}_{l}|N^{2})$ & $0$&$O(T|\mathcal{D}_{l}|N^{2})$ \\							
			\bottomrule
		\end{tabular}
	\end{minipage}
\end{table}
\end{footnotesize}

Fig.\,\ref{Fig:nonscalable} compares the achievable BER of the proposed distributed EP detector with other detectors investigated in \cite{decentalized_MMSE}.  In this case, the simulation parameters are $L=16$, $N=8$, and $K=16$. The SNR is defined as $\mathrm{SNR} = 1/\sigma^{2}$. The results are obtained using Monte Carlo simulation with 10,000 independent channel realizations. We denote ``deEP'' as the distributed EP detector. It can be observed that the performance of the proposed distributed EP detector is comparable with that of the Level 1 detector with only one EP iteration and outperforms the centralized Level 4 detector with $T = 5$ iterations.

We also compare the performance of the distributed EP detector with other receivers in scalable cell-free massive MIMO systems. Fig.\,\ref{Fig:DCC} shows that the distributed EP detector outperforms both the centralized and distributed MMSE detectors. The simulation parameters are the same as those in Fig.\,\ref{Fig:nonscalable}. The performance loss is acceptable when compared to conventional cell-free massive MIMO and the computational complexity is significantly decreased from $O(KN^{2})$ to $O(|\mathcal{D}_{l}|N^{2})$. Furthermore, we compare the complexity and  fronthaul overhead of distributed EP detector with other receivers for cell-free massive MIMO systems in Tables \ref{tbl:complexity} and \ref{tbl:fronthaul}, respectively, where $T$ is the number of iterations in distributed EP detector.
\begin{footnotesize}
\begin{table}[t]
	\centering
	\renewcommand{\arraystretch}{1.5}
	\begin{minipage}[c]{1\columnwidth}
		\centering
		\caption{.~~Fronthaul overhead}
		\label{tbl:fronthaul}
		\begin{tabular}{@{}lcc@{}}
		  \toprule
		   Detectors &  Statistical parameters & Coherence block   \\
		  \midrule
		  Level 1  &$0$&$0$  \\	
		  \midrule
		  Level 4  &$KLN^{2}/2$ &$\tau_{c}NL$  \\
		  \midrule
		  deEP  &$0$  &$\sum\limits_{l=1}^{L}(\tau_{c}-\tau_{p})2T(|\mathcal{D}_{l}|+1)$ \\			
		 \bottomrule
		\end{tabular}
	\end{minipage}
\end{table}
\end{footnotesize}
\subsection{Downlink Precoding}\label{sec:down_link}
The downlink precoding for cell-free massive MIMO can be classified into centralized and distributed manner, respectively. Consider a downlink cell-free massive MIMO network with single-antenna users. The downlink data transmitted by the $l$-th AP to all users is given by
\begin{equation}\label{eqdata}
  {\bf x}_{l} = \sum_{k=1}^{K} \sqrt{\rho_{l,k}}{\bf w}_{l,k}q_{k}+{\bf n},
\end{equation}
where ${\bf w}_{l,k}$ is the precoding vector between the $l$-th AP and the $k$-th user with ${\bf E}\{\|{\bf w}_{l,k}\|^{2}\}=1$, and $\rho_{l,k}$ is the normalized transmit power satisfying the per-AP power constraint. We assume that the data symbol $q_{k}$ has unit power, zero mean, and is uncorrelated. The goal of downlink precoding is to design precoder ${\bf w}_{l,k}$ for each user to achieve a some performance objective.

Early works investigated the  maximum ratio transmission (MRT) beamforming to maintain system scalability and  cope with single-antenna APs \cite{cell_free}. Then, a modified MRT precoding was proposed to eliminate the self-interference without action required at the receivers \cite{Modified_MRT}, but it requires CSI exchange among the APs which is not scalable when the numbers of users and APs is large. In contrast, the enhanced MRT precoding scheme was proposed in \cite{Enhanced_MRT}, in which each AP only needs its own channel estimates to construct the precoding vectors. The enhanced MRT precoding can greatly boost the channel hardening, enabling the users to reliably decode data based on statistical CSI.

Subsequently, centralized zero-forcing (ZF) processing was proposed without pilot contamination effect \cite{cell_free}. However, the centralized ZF precoding requires the CPU to obtain the instantaneous CSI from all  APs to design the precoding vectors and is not scalable when the number of APs are users is large. A local full-pilot ZF scheme was then proposed, where each AP uses its own local channel estimates to design a ZF precoder. It suppress own interference while the centralized ZF suppresses the interference from other APs. Afterward, two fully distributed precoding approaches, named the local partial ZF and local protective partial ZF, were proposed. The advantage is that they can provide interference cancelation gain without additional fronthaul overhead and can be implemented with very few number of antenna. Based on the two aforementioned precoding approaches, closed-form expressions for achievable downlink SE were derived which takes channel estimation errors and pilot contamination effect into consideration. In \cite{Team_MMSE}, a novel distributed precoding design, which generalizes the classical centralized MMSE precoding to distributed processing in cell-free massive MIMO, was proposed. Based on the {\em theory of teams}, a set of necessary and sufficient conditions for optimal TMMSE precoding was derived.
The proposed team MMSE precoding can improve upon previous local precoding based on local CSI only. Furthermore, it can also be efficiently
implemented in a sequential fashion, which is an idea that has been explored already in \cite{MMSE_optimal}.

The above-aforementioned works investigate the precoder design for single-antenna user. For multi-antenna user, a distributed processing for cooperative precoding was proposed \cite{OTA_precoding}. It adopts a over-the-air (OTA) signal mechanism to allow each AP to exchange CSI, which enjoys desirable flexibility and scalability properties because OTA signaling does not scale with the number of APs or users.

\subsection{Power Control and Allocation}\label{sec:power_control}
For cell-free massive MIMO systems, one of the significant challenges is power control and allocation. In general, the procedure of controlling the uplink transmit powers is {\em power control} while the procedure of allocating the downlink transmit power is {\em power allocation}. The power control/allocation coefficients should be selected to maximize a given performance objective and the objective. The objective may be the max-min rate, sum rate, and energy efficiency. Based on these objectives, several power control approaches have been developed such as Max-Min fairness power control, power control with user prioritization, and power control with AP selection.

\begin{itemize}
  \item \textbf{Optimization-based methods}: The max-min-based power control method was proposed in \cite{cell_free}, where it is shown that a max-min solution for the user transmit power yields a highly equalized performance across the network and the channel conditions will determine the service quality.  The max-min fairness problem is regarded as convex or quasi-convex problem and the optimal solution is obtained by utilizing bisection search and convex optimization, geometric programming, or second-order cone program algorithms \cite{cell_free,Nayebi_precoding}. Furthermore, some papers considered the max-min fairness problem as non-convex and decomposed the  non-convex problem into several subproblems \cite{precoding_JSFD}, thereby some convex optimization algorithms can be applied. However, the overall system performance may be reduced by the users with bad channel conditions if max-min fairness is considered. Therefore, the maximization of the sum SE which represents the overall SE performance of the network was investigated \cite{SCA1,SCA2}. However, these aforementioned optimization-based algorithms do not represent a convenient option because they must be found in an iterative and centralized fashion, meaning that they are not distributed and is not scalable. Therefore, many researchers considered alternative schemes such as deep learning (DL) and Heuristic algorithms.

  \item \textbf{DL-based methods}: Recently, DL has been shown an effective tool for solving power control problems in wireless communications \cite{powercontrol4, DL_Power, JSAC_Power}. Therefore, some works exploit DL for power control in cell-free massive MIMO systems. In \cite{DL_Power_Con}, a max sum SE problem in an uplink cell-free massive MIMO system was studied by using artificial neural networks (ANNs), in which the UE positions were taken as input and the power control policy as output. Then, a deep convolutional neural network (DCNN) was considered in an uplink cell-free mMIMO system with limited-fronthaul, where
      the large-scale fading information was exploited to predict the max sum SE power control policy \cite{DL_Power_JSAC}.

  \item \textbf{Heuristic algorithms}:    Fractional power control is a classical heuristic algorithm in cellular networks \cite{power95JSAC}. The principle of fractional power control is controlling the user transmit power to compensate for a fraction of the path-loss among.  It can minimize the variance of the large-scale SINR distribution and the final solution is $p_{k}\propto \sqrt{\beta_{k}}$ where $\beta_{k}$ is the large-scale gain. In the cell-free massive MIMO, the fractional power control is generalized as
      \begin{equation}\label{eqfpc}
      p_{k}  \propto \frac{1}{(\sum_{l}^{L}\beta_{kl})^{v}}.
      \end{equation}
      where $v=[0,1]$.
\end{itemize}

\section{Ongoing and Future Research Problems}\label{sec:inter}
Although cell-free massive MIMO has been investigated for several years, some challenges are still required to be solved. Furthermore, several new technologies have been proposed as enabling technologies for wireless communications. Therefore, many researchers aim to incorporate these technologies into cell-free massive MIMO. In this section, we will elaborate on several opportunities and issues for cell-free massive MIMO, including several emerging technologies and practical issues.
\subsection{Cell-Free Massive MIMO with Reconfigurable Intelligent Surfaces}
RIS is capable of significantly increasing the SE to meeting the demanding requirements for 6G wireless communication networks. A RIS with many reflective elements is designed to adjust the reflected beam in different directions \cite{RIS_Huang,RIS_Wu,RIS_Tang, RIS_Yu}. Recently, the uplink performance of a RIS-aided cell-free massive MIMO system was analyzed \cite{Cell_free_RIS1}. An alternative optimization algorithm was then applied to decouple the max-min rate optimization problem into phase shift design and power allocation problems. The power allocation problem was solved by using a standard geometric programming algorithm while  semidefinite programming was utilized to design the phase shifts. Analytical and simulation results showed the performance of the cell-free massive MIMO systems can be improved by RIS \cite{Cell_free_RIS1}. Subsequently,  an aggregated channel estimation approach was proposed to reduce the overhead required for channel estimation by utilizing sufficient information for data processing in \cite{Cell_free_RIS2}. It was shown that using RISs in cell-free massive MIMO systems bring performance benefits, especially if the direct links between the APs and the users are of insufficient quality with a high probability.
\subsection{Machine Learning for Cell-Free Massive MIMO}
Recently, machine learning (ML) has been applied to physical layer communications \cite{DL_WC}, such as CSI feedback \cite{DL_CSI}, channel estimation \cite{DL_CE}, MIMO detection \cite{DL_Det}, and power control \cite{DL_Power,JSAC_Power}. Therefore, some researchers considered using the ML tool for improving the system performance of cell-free massive MIMO. In \cite{DL_Cascade}, a cascade of two deep neural networks was proposed to calibrate the TDD reciprocity in cell-free massive MIMO systems. It is scalable and avoids the need of antenna cooperation for reciprocity calibration. For DL-based power control, a deep convolutional neural network (DCNN) combined with large-scale-fading (LSF) was employed to determine a mapping between the LSF coefficients and the optimal power and achieved great performance \cite{DL_Power_Con}.
\subsection{Cell-Free Massive MIMO with Federated Learning}
DL has developed dramatically thanks to the availability of the large number of data. In general, the data samples are often acquired on edge equipments, such as smart phones, vehicles, and sensors. Owing to the privacy considerations, data cannot be shared, which has aroused wide investigation on distributed learning with centralized aggregations. Thus, an emerging paradigm for enabling learning at the edge, named \emph{federated learning}, was proposed. In \cite{Cell_free_FL}, a scheme for cell-free massive MIMO networks to support the FL framework was proposed where the APs were considered as relays to transmit the training updates between the CPU and users. Then, a mixed-timescale stochastic nonconvex optimization problem was formulated to minimize the time of one FL process. Simulation results have shown that the proposed algorithms can reduce the training time significantly.
\subsection{Supporting IoT with Cell-Free Massive MIMO}
IoT is an emerging paradigm for future communication systems, where a large number of machine type devices transmit small data packets to a base station occasionally. It has been investigated for co-located massive MIMO systems \cite{IoT18Liu,IoT19Liu} as massive MIMO can support massive connectivity due to its high capacity, reliability, and energy efficiency. Recently,  cell-free massive MIMO was investigated as a way to support IoT systems. By leveraging the macro-diversity gain and better coverage resulting from distributed antennas, cell-free massive MIMO can outperform the IoT network supported by co-located massive MIMO systems. Therefore, it is interesting to consider and further study the use of cell-free massive MIMO to support IoT.
\subsection{Advanced Distributed Signal Processing}
As mentioned in Section \ref{sec:intro}, it is critically important to design a low cost and scalable algorithm to achieve the potential performance. Although several distributed signal processing algorithms have been investigated for the receiver and precoder design \cite{decentalized_MMSE, OTA_precoding}, the performance is far from the centralized algorithms. Therefore, we need to design  cell-free systems with  advanced distributed signal processing to improve the system performance, scalability, and robustness, especially in transmitter design and power control.
\subsection{Fronthaul-Limited}
In cell-free massive MIMO system, additional overhead is required for exchanging information between the APs and CPU. However, the fronthaul is limited in practical systems. In order to reduce the fronthaul load, quantizing the transmitted signals and structured lattice codes were utilized \cite{quantization1,DL_fronthaul,quantization2,quantization3,quantization4}. How to design the signal processing algorithm with fronthaul-limited capacity should be investigated in the future.
\subsection{Practical Issues}
To achieve ubiquitous cell-free massive MIMO, a low-cost and low-complexity deployment scheme is required. Therefore, the network architecture based on a novel hardware platform should be considered. Recently, an appropriate and cost-effective architecture, named {\em radio strip system}, was proposed \cite{MMSE_optimal}. It can reduce the fronthaul and deployment cost. Therefore, it is promising to design the signal processing algorithms based on the {\em radio strip system} to achieve ubiquitous cell-free massive MIMO.
\section{Conclusion}\label{sec:con}
This article presented a comprehensive study and review of the signal processing algorithms in cell-free massive MIMO systems. Several recent contributions were highlighted, including user association and pilot assignment, channel estimation, uplink combining and downlink precoding, as well as power control and allocation. Furthermore, some emerging technologies such as machine learning, RIS and federated learning, which can improve the cell-free massive MIMO, were introduced. Given its impressive performance and uniform connectivity for all users, cell-free massive MIMO  inherits the superiority of  distributed antennas and massive MIMO systems and is thus a  very promising enabling technology for 6G wireless communication networks.


\begin{thebibliography}{99}

\bibitem{massiveMIMO}
T. L. Marzetta, ``Noncooperative cellular wireless with unlimited numbers
of base station antennas,'' {\em IEEE Trans. Wireless Commun.}, vol. 9,
no. 11, pp. 3590-3600, Nov. 2010.

\bibitem{6G_khaled}
K. B. Letaief, W. Chen, Y. Shi, J. Zhang, and Y.-J.-A. Zhang, ``The
roadmap to 6G: AI empowered wireless networks,'' {\em IEEE Commun.
Mag.}, vol. 57, no. 8, pp. 84-90, Aug. 2019.

\bibitem{6G_matthaiou}
M. Matthaiou, O. Yurduseven, H. Q. Ngo, D. Morales-Jimenez, S. L. Cotton, and V. F. Fusco, ``The road to 6G: Ten physical layer challenges for communications engineers,'' {\em IEEE Commun. Mag.}, vol. 59, no. 1, pp. 64-69, Jan. 2021.

\bibitem{6G_you}
X. You et al., ``Towards 6G wireless communication networks: Vision,
enabling technologies, and new paradigm shifts,'' {\em Sci. China Inf. Sci.},
vol. 64, no. 1, pp. 1-74, 2021.

\bibitem{beyond5G}
J. Zhang, E. Bj\"{o}rnson, M. Matthaiou, D. W. K. Ng, H. Yang, and D. J. Love, ``Prospective multiple antenna technologies for beyond 5g,'' {\em IEEE J. Sel. Areas Commun.}, vol. 38, no. 8, pp. 1637-1660, Aug. 2020.

\bibitem{DAS}
W. Choi, J. G. Andrews, ``Downlink performance and capacity of distributed antenna systems in a multicell environment,'' {\em IEEE Trans. Wireless Commun.}, vol. 6, no. 1, pp. 69-73, Jan. 2007.

\bibitem{Zhang}
J. Zhang, R. Chen, J. G. Andrews, A. Ghosh, and R. W. Heath Jr, ``Networked MIMO with clustered linear precoding,'' {\em IEEE Trans. Wireless Commun.}, vol. 8, no. 4, pp. 1910-1921, Apr. 2009.

\bibitem{Network}
H. Huh, A. M. Tulino, and G. Caire, ``Network MIMO with linear zero-forcing beamforming: Large system analysis,
impact of channel estimation, and reduced-complexity scheduling,'' {\em IEEE Trans. Inf. Theory}, vol. 58, pp. 2911-2934, May 2012.

\bibitem{CoMP}
3GPP, Coordinated multi-point operation for LTE physical layer aspects, 2013. http://www.3gpp.org/ftp/Specs/archive/
36 series/36.819/36819-b20.zip.

\bibitem{cell_free}
H. Q. Ngo, A. Ashikhmin, H. Yang, E. G. Larsson, and T. L. Marzetta,
''Cell-free massive MIMO versus small cells,'' {\em IEEE Trans. Wireless
Commun.}, vol. 16, no. 3, pp. 1834-1850, Mar. 2017.

\bibitem{Ubiquitous}
G. Interdonato, E. Bj\"{o}rnson, H. Q. Ngo, P. Frenger, and E. G. Larsson, ``Ubiquitous cell-free massive MIMO
communications,'' {\em EURASIP J. Wireless Commun. and Netw.}, vol. 2019, no. 1, pp. 197-206, 2019.

\bibitem{FDD_cell}
A. Abdallah and M. M. Mansour, ``Efficient angle-domain processing for FDD-based cell-free massive MIMO systems''
{\em IEEE Trans. Commun.}, vol. 68, no. 4, 2188-2203, Apr. 2020.

\bibitem{FDD_cell_free}
S. Kim and B. Shim, ``FDD-based cell-free massive MIMO systems,''
in {\em Proc. IEEE Int. Signal Process. Adv. Wireless Commun. Workshop
(SPAWC)}, Kalamata, Greece, Jun. 2018, pp. 1-5.

\bibitem{DCC}
E. Bj\"{o}rnson, N. Jald\'{e}n, M. Bengtsson, and B. Ottersten, ``Optimality
properties, distributed strategies, and measurement-based evaluation
of coordinated multicell OFDMA transmission,'' {\em IEEE Trans. Signal
Process.}, vol. 59, no. 12, pp. 6086-6101, Dec. 2011.

\bibitem{scalable}
E. Bj\"{o}rnson and L. Sanguinetti, ``Scalable cell-free massive MIMO
systems,'' {\em IEEE Trans. Commun.}, vol. 68, no. 7, pp.
4247-4261, Jul. 2020.

\bibitem{performance_analysis1}
\"{O}. \"{O}zdogan, E. Bj\"{o}rnson, and J. Zhang, ``Performance of cell-free massive MIMO with rician fading and phase shifts,'' {\em IEEE Trans. Wireless Commun.}, vol. 18, no. 11, pp. 5299-5315, Nov. 2019.

\bibitem{performance_analysis2}
Z. Wang, J. Zhang, E. Bj\"{o}rnson, and B. Ai, ``Uplink performance of cell-free massive MIMO over spatially correlated Rician fading channels,'' {\em IEEE Commun. Lett.}, vol. 25, no. 4, pp. 1348-1352, Apr. 2020.

\bibitem{performance_analysis3}
S.-N. Jin, D.-W. Yue, and H. H. Nguyen, ``Spectral and energy efficiency in cell-free massive MIMO systems over
correlated Rician fading,'' {\em IEEE Syst. J.}, pp. 1-12, May. 2020

\bibitem{performance_analysis6}
H. Q. Ngo, L. Tran, T. Q. Duong, M. Matthaiou, and E. G. Larsson, ``On the total energy efficiency of cell-free massive MIMO,'' {\em IEEE Trans. Green Commun. Netw}, vol. 2, no. 1, pp. 25-39, 2018.

\bibitem{performance_analysis8}
H. V. Nguyen, V. D. Nguyen, O. A. Dobre, S. K. Sharma, S. Chatzinotas, B. Ottersten, and O. S. Shin, ``On the spectral and energy efficiencies of full-duplex cell-free massive MIMO,'' {\em IEEE J. Sel. Areas Commun.}, vol. 38, no. 8, pp. 1698-1718, Aug. 2020.

\bibitem{cell_free_Zhang}
J. Zhang, S. Chen, Y. Lin, J. Zheng, B. Ai, and L. Hanzo, ``Cell-free massive MIMO: A new next-generation paradigm,'' {\em IEEE Access}, vol. 7, pp. 99878-99888, Sep. 2019.

\bibitem{survey2}
S. Chen, J. Zhang and B. Ai, ``A survey on user-centric cell-free massive MIMO systems,'' {\em arXiv
preprint arXiv:2104.13667}, 2021.

\bibitem{survey3}
H. A. Ammar, R. Adve, S. Shahbazpanahi, G. Boudreau, and K. V. Srinivas, ``User-centric cell-free massive MIMO networks: A survey of opportunities, challenges and solutions,'' {\em arXiv preprint arXiv:2104.14589}, 2021.

\bibitem{random_pilot}
M. Attarifar, A. Abbasfar, and A. Lozano, ``Random vs structured pilot
assignment in cell-free massive MIMO wireless networks,'' in {\em Proc. IEEE Int. Conf. Commun. Workshops (ICC Workshops)}, Kansas City, MO, USA, May 2018, pp. 1-6.


\bibitem{local_based}
Y. Zhang, H. Cao, P. Zhong, C. Qi, and L. Yang, ``Location-based greedy
pilot assignment for cell-free massive MIMO systems'' in {\em Proc. IEEE Int.
Conf. Comput. Creativity}, Chengdu, China, 2018, pp. 392-396.

\bibitem{graph_coloring}
H. Liu, J. Zhang, S. Jin, and B. Ai, ``Graph coloring based pilot
assignment for cell-free massive MIMO systems,'' {\em IEEE Trans. Veh.
Technol.}, vol. 69, no. 8, pp. 9180-9184, Aug. 2020.


\bibitem{Pilot_Hungarian}
S. Buzzi, C. D'Andrea, M. Fresia, Y. Zhang, S. Feng, ``Pilot Assignment
in Cell-Free Massive MIMO based on the Hungarian Algorithm,'' {\em IEEE
Wireless Commun. Lett.}, vol. 10, no. 1, 34-37, Jan. 2021.


\bibitem{Pilot_GNN}
W. Zeng, Y. He, B. Li, and S. Wang, ``Pilot assignment for cell-free massive MIMO systems using a weighted graphic
framework,''” {\em IEEE Trans. on Veh. Technol.}, pp. 1-1, 2021.

\bibitem{decentalized_MMSE}
E. Bj\"{o}rnson and L. Sanguinetti, ``Making cell-free massive MIMO competitive with MMSE processing and centralized
implementation,'' {\em IEEE Trans. Wireless Commun.}, vol. 19, no. 1, pp. 77-90, Jan. 2020.

\bibitem{FFDet}
Y. Jin, J. Zhang, S. Jin, and B. Ai, ``Channel estimation for cell-free
mmWave massive MIMO through deep learning,'' {\em IEEE Trans. Veh. Technol.}, vol. 68, no. 10, pp. 10325-10329, Oct. 2019.

\bibitem{MMSE_optimal}
Z. H. Shaik, E. Bj\"{o}rnson, and E. G. Larsson, ``MMSE-optimal sequential processing for cell-free massive MIMO with ¨
radio stripes,'' {\em arXiv preprint arXiv:2012.13928}, 2020.

\bibitem{PM}
E. G. Larsson and J. Jalden, ``Fixed-complexity soft MIMO detection
via partial marginalization,'' {\em IEEE Trans. Signal Process.},
vol. 56, no. 8, pp. 3397-3407, Aug. 2008.

\bibitem{EP_distributed2}
H. He, H. Wang, X. Yu, J. Zhang, S.H. Song, and K. B. Letaief, ``Distributed expectation propagation detector for cell-free massive MIMO,'' {\em IEEE Global Commun. Conf. (GLOBECOM)}, Madrid, Spain, Dec. 2021.

\bibitem{JCD}
H. Song, X. You, C. Zhang, O. Tirkkonen, C. Studer, ``Minimizing pilot overhead in cell-Free massive MIMO systems via joint estimation and detection,'' in {\em Proc. IEEE 21st Int. Workshop Signal Process. Adv. Wireless Commun.(SPAWC)}, Atlanta, GA, USA, May. 2020, pp. 1-5.

\bibitem{Enhanced_MRT}
G. Interdonato, H. Q. Ngo, and E. G. Larsson, ``Enhanced normalized conjugate beamforming for cell-free massive MIMO,'' {\em IEEE Trans.
Commun.}, vol. 69,  no. 5, pp. 2863-2877, May. 2021.

\bibitem{JMRZF}
L. Du, L. Li, H. Q. Ngo, T. C. Mai, and M. Matthaiou, ``Cell-free massive MIMO: Joint maximum-ratio and zero-forcing precoder with power control'' {\em IEEE Trans. Commun.}, vol. 69, no. 6, 3741-3756, Jun. 2021.

\bibitem{Local_ZF}
G. Interdonato, M. Karlsson, E. Bj\"{o}rnson, and E. G. Larsson, ``Local partial zero-forcing precoding for cell-free massive MIMO'' {\em IEEE Trans. Wireless Commun.}, vol. 19, no. 7, pp. 4758-4774, Jul. 2020.

\bibitem{Team_MMSE}
L. Miretti, Emil Bj\"{o}rnson, and David Gesbert, ``Team MMSE precoding with applications to
cell-free massive MIMO,'' {\em arXiv preprint arXiv:2104.15027}, 2020.

\bibitem{OTA_precoding}
I. Atzeni, B. Gouda, and A. Tölli, ``Distributed precoding design via over-the-air signaling for cell-free massive
MIMO,'' {\em IEEE Trans. Wireless Commun.}, vol. 20, no. 2, pp. 1201-1216, Feb. 2021.

\bibitem{SCA1}
T. H. Nguyen, T. K. Nguyen, H. D. Han et al., ``Optimal power control and load balancing for uplink cell-free multi-user massive MIMO,'' {\em IEEE Access}, vol. 6, pp. 14462-14473, 2018.

\bibitem{SCA2}
J. Francis, P. Baracca, S. Wesemann, and G. Fettweis, ``Downlink power control in cell-free massive MIMO with partially distributed access points,'' in {\em Proc. IEEE VTC2019-Fall}, Honolulu, HI, USA, Sep. 2019, pp. 1-7.

\bibitem{power_control2}
Y. Al-Eryani, M. Akrout, and E. Hossain, ``Multiple access in cell-free
networks: Outage performance, dynamic clustering, and deep reinforcement learning-based design,'' {\em IEEE J. Sel. Areas Commun.}, vol. 39, no. 4, pp. 1028-1042, Apr. 2021.

\bibitem{DL_Power_Con}
C. D'Andrea, A. Zappone, S. Buzzi, and M. Debbah, ``Uplink power control in cell-free massive MIMO via deep learning,'' in {\em Proc. IEEE CAMSAP}, Le gosier, Guadeloupe, 2019, pp. 554-558.

\bibitem{DL_Power_JSAC}
M. Bashar, A. Akbari, K. Cumanan, H.-Q. Ngo, A. G. Burr, P. Xiao, M. Debbah, and J. Kittler, ``Exploiting deep learning in limited-fronthaul cell-free massive MIMO uplink, {\em IEEE J. Sel. Areas in Commun.}, vol. 38, no. 8, pp. 1607-1678, Jun. 2020.

\bibitem{power_control3}
R. Nikbakht and A. Lozano, ``Uplink fractional power control for cell-free wireless networks,'' in {\em Proc. IEEE ICC}, Shanghai, China, May. 2019, pp. 1-5.


\bibitem{Cell_free_RIS1}
M. Bashar, K. Cumanan, A. G. Burr, P. Xiao, and M. Di Renzo, ``On the performance of reconfigurable intelligent surface-aided cell-free massive mimo uplink,'' in {\em Proc. IEEE GLOBECOM}, Taipei, Taiwan, 2020, pp. 1-6.

\bibitem{Cell_free_RIS2}
T. V. Chien, H. Q. Ngo, S. Chatzinotas, M. D. Renzo, and
B. Ottersten, ``Reconfigurable intelligent surface-assisted Cell-Free
Massive MIMO systems over spatially-correlated channels,'' {\em arXiv preprint arXiv:2104.08648}, 2020.

\bibitem{DL_Cascade}
N. Athreya, V. Raj, and S. Kalyani, ``Beyond 5G: Leveraging cell free TDD massive MIMO using cascaded deep learning,'' {\em IEEE Wireless Commun. Lett.}, vol. 9, no. 9, pp. 1533-1537, May. 2020.

\bibitem{Cell_free_FL}
T. T. Vu, D. T. Ngo, N. H. Tran, H. Q. Ngo, M. N. Dao, and R. H. Middleton, ``Cell-free massive MIMO for wireless federated learning,'' {\em IEEE Trans. Wireless Commun.}, vol. 19, no. 10, pp. 6377-6392, Oct. 2020.

\bibitem{quantization1}
M. Bashar, K. Cumanan, A. G. Burr, H. Q. Ngo, M. Debbah, and P. Xiao, ``Max–min rate of cell-free massive MIMO
uplink with optimal uniform quantization,'' {\em IEEE Trans. Commun.}, vol. 67, no. 10, pp. 6796-6815, Jul. 2019.

\bibitem{DL_fronthaul}
M. Bashar, A. Akbari, K. Cumanan, H. Q. Ngo, A. G. Burr, P. Xiao, M. Debbah, and J. Kittler, ``Exploiting deep learning in limited-fronthaul cell-free massive MIMO uplink,'' {\em IEEE J. Sel. Areas Commun.}, vol. 38, no. 8,pp. 1678-1697, Jun. 2020.

\bibitem{quantization2}
D. Maryopi, M. Bashar, and A. Burr, ``On the uplink throughput of zero forcing in cell-free massive MIMO with coarse
quantization,'' {\em IEEE Trans. Veh. Technol.}, vol. 68, no. 7, pp. 7220-7224, Jun. 2019.

\bibitem{quantization3}
M. Bashar, K. Cumanan, A. G. Burr, H. Q. Ngo, E. G. Larsson, and P. Xiao, ``Energy efficiency of the cell-free massive MIMO uplink with optimal uniform quantization,'' {\em IEEE Trans. Green Commun. Netw,}, vol. 3, no. 4, pp. 971-987, Jul. 2019.

\bibitem{quantization4}
M. Bashar et al., ``Uplink spectral and energy efficiency of cell-free massive MIMO with optimal uniform quantization,'' {\em IEEE Trans. Commun.}, vol. 69, no. 1, pp. 223-245, Jan. 2021.

\bibitem{user_association}
Carmen D'Andrea, Erik G. Larsson, ``User Association in Scalable Cell-Free Massive MIMO Systems''
\emph{arxiv preprint} \emph{arXiv:2103.05321}, 2021.

\bibitem{Superimposed}
Y. Zhang, X. Qiao, et al., ``Superimposed pilots are beneficial for mitigating pilot contamination in cell-free massive
mimo,'' {\em IEEE Commun. Lett.}, vol. 25, no. 1, pp. 279-283, Jan. 2020.

\bibitem{DP_channel_estimation}
J. Xu, X. Wang, P. Zhu, and X. You, ``Privacy-preserving channel estimation in cell-Free hybrid massive MIMO systems,'' {\em IEEE Trans. Wireless Commun.}, vol. 20, no. 6, pp. 3815-3830, Jun. 2021.

\bibitem{LSFD}
E. Nayebi, A. Ashikhmin, T. L. Marzetta, and B. D. Rao, ``Performance of cell-free massive MIMO systems with MMSE and LSFD receivers,'' in {\em Proc. 50th Asilomar Conf. Signals, Syst. Comput.}, Pacific Grove, CA, USA, Nov. 2016, pp. 203-207.

\bibitem{soft_detection}
Carmen D'Andrea, E. Bjornson, and E. G. Larsson, ``Improving cell-free massive MIMO by local per-bit soft detection,'' {\em arXiv preprint arXiv:2012.13928}, 2020.

\bibitem{EP_principle}
T. P. Minka, ``A family of algorithms for approximate Bayesian Inference,'' Ph.D. dissertation, Dept. Elect. Eng. Comput. Sci., MIT, Cambridge, MA, USA, 2001.

\bibitem{Decentralized-AMP}
C. Jeon, K. Li, J. R. Cavallaro, and C. Studer, ``Decentralized equalization with feedforward architectures for massive MU-MIMO,'' {\em IEEE Trans. Signal Process.}, vol. 67, no. 17, pp. 4418-4432, Sep. 2019.

\bibitem{EP_distributed}
H. Wang, A. Kosasih, C. Wen, S. Jin, and W. Hardjawana, ``Expectation propagation detector for extra-large scale massive MIMO,'' {\em IEEE Trans.
Wireless Commun.}, vol. 19, no. 3, pp. 2036-2051, Mar. 2020.

\bibitem{Takeuchi}
K. Takeuchi, ``Rigorous dynamics of expectation-propagation-based signal recovery from unitarily invariant measurements,'' {\em IEEE Trans. Inf. Theory.}, vol. 66, no. 1, 368-386, Oct. 2019.

\bibitem{EP_detector}
J. C\'{e}spedes, P. M. Olmos, M. Sánchez-Fernández, and F. Perez-Cruz,
``Expectation propagation detection for high-order high-dimensional
MIMO systems,'' {\em IEEE Trans. Commun.}, vol. 62, no. 8, pp. 2840-2849,
Aug. 2014.

\bibitem{HeJSTSP}
H. He, C.-K. Wen, and S. Jin, ``Bayesian optimal data detector for hybrid
mmWave MIMO-OFDM systems with low-resolution ADCs,'' {\em IEEE J. Sel. Topics Signal Process.}, vol. 12, no. 3, pp. 469-483, Jun. 2018.

\bibitem{Modified_MRT}
M. Attarifar, A. Abbasfar, and A. Lozano, ``Modified conjugate beambforming for cell-free massive MIMO,'' {\em IEEE Wireless Commun. Lett.}, vol. 8, no. 2, pp. 616-619, Apr. 2019.

\bibitem{Nayebi_precoding}
E. Nayebi, A. Ashikhmin, T. L. Marzetta, H. Yang, and B. D. Rao, ``Precoding and power optimization in cell-free
massive MIMO systems,'' {\em IEEE Trans. Wireless Commun.}, vol. 16, no. 7, pp. 4445-4459, Jul. 2017.

\bibitem{precoding_JSFD}
\"{O}. T. Demir and E. Bj\"{o}rnson, ``Joint power control and LSFD for wireless-powered cell-free massive MIMO,''
 {\em IEEE Trans. Wireless Commun.}, vol. 20, no. 3, pp. 1756-1769, 2021.

\bibitem{powercontrol4}
F. Liang, C. Shen, W. Yu, and F. Wu, ``Towards optimal ower control via ensembling deep neural networks,'' {\em IEEE Trans. Commun.}, vol. 68, no. 3, pp. 1760-1776, Mar. 2019.

\bibitem{DL_Power}
H. Sun, X. Chen, Q. Shi, M. Hong, X. Fu, and N. D. Sidiropoulos, ``Learning to optimize: Training deep neural networks for interference management,'' {\em IEEE Trans. Signal Process., vol.} 66, no. 20, pp. 5438-5453, Oct. 2018.

\bibitem{JSAC_Power}
H. Lee, S. H. Lee, and T. Q. S. Quek, ``Deep Learning for Distributed Optimization: Applications to Wireless Resource Management," {\em IEEE J. Sel. Areas Commun.}, Vol. 37, No. 10, pp. 2251-2266, Oct. 2021.

\bibitem{power95JSAC}
R. D. Yates, ``A framework for uplink power control in cellular radio systems," {\em IEEE J. Sel. Areas Commun.}, vol. 13, no. 7, pp. 1341-1347,
Sep. 1995.

\bibitem{RIS_Huang}
C. Huang, A. Zappone, G. C. Alexandropoulos, M. Debbah, and C. Yuen, ``Reconfigurable intelligent surfaces for energy efficiency in wireless communication'' {\em IEEE Trans. Wireless Commun.}, vol. 18, no. 8, pp. 4157-4170, Aug. 2019.

\bibitem{RIS_Wu}
Q. Wu and R. Zhang, ``Intelligent reflecting surface enhanced wireless
network via joint active and passive beamforming,'' {\em IEEE Trans. Wireless
Commun.}, vol. 18, no. 11, pp. 5394-5409, Nov. 2019.

\bibitem{RIS_Tang}
W. Tang et al., ``Wireless communications with reconfigurable intelligent
surface: Path loss modeling and experimental measurement,'' {\em IEEE
Trans. Wireless Commun.}, vol. 20, no. 1, pp. 421-439, Jan. 2021.

\bibitem{RIS_Yu}
X. Yu, D. Xu, Y. Sun, D. W. K. Ng, and R. Schober, ``Robust and secure
wireless communications via intelligent reflecting surfaces,'' {\em IEEE J.
Sel. Areas Commun.}, vol. 38, no. 11, pp. 2637-2652, Jul. 2020.

\bibitem{DL_WC}
H. He, S. Jin, C.-K. Wen, F. Gao, G. Y. Li, and Z. Xu, ``Model-driven deep learning for physical layer communications,'' {\em IEEE Wireless Commun.}, vol. 26, no. 5, pp. 77-83, 2019.

\bibitem{DL_CSI}
C. K. Wen, W. T. Shih, and S. Jin, ``Deep Learning for Massive MIMO CSI Feedback,'' {\em IEEE Wireless Commun. Lett.}, vol. 7, no. 5, pp. 748-751, Oct. 2018.

\bibitem{DL_CE}
H. He, C.-K. Wen, S. Jin, and G. Y. Li, ``Deep learning-based channel estimation for beamspace mmWave massive MIMO systems,'' {\em IEEE Wireless Commun. Lett.}, vol. 7, no. 5, pp. 852-855, Oct. 2018.

\bibitem{DL_Det}
H. He, C.-K. Wen, S. Jin, and G. Y. Li, ``Model-driven deep learning for MIMO detection,'' {\em IEEE Trans. Signal Process.}, vol. 68, pp. 1702-1715, Feb. 2020.

\bibitem{IoT18Liu}
L. Liu, E. G. Larsson, W. Yu, P. Popovski, \v{C}. Stefanovi\'{c}, and E. de Carvalho, ``Sparse signal processing for grant-free massive connectivity:
A future paradigm for random access protocols in the Internet of Things,'' {\em IEEE Signal Process. Mag.}, vol. 35, no. 5, pp. 88–99, Sep. 2018.

\bibitem{IoT19Liu}
L. Liu and W. Yu, ``Massive connectivity with massive MIMO-Part I: Device activity detection and channel estimation,'' {\em IEEE Trans. Signal
Process.}, vol. 66, no. 11, pp. 2933–2946, Jun. 2018.


%
%
%
%

\end{thebibliography}
\end{document}